\documentclass[english,aps,prb,showpacs,floats,twocolumn]{revtex4}
\usepackage[latin9]{inputenc}
\usepackage{babel}

\usepackage{graphicx}

\usepackage[unicode=true, pdfusetitle, bookmarks=true,bookmarksnumbered=false,bookmarksopen=false, breaklinks=false,pdfborder={0 0 1},backref=page,colorlinks=false]{hyperref}

\usepackage{amssymb}
\usepackage{amsmath}

\begin{document}

\title{Magnetization reversal via internal spin waves in magnetic nanoparticles}

\author{D. A. Garanin$^{1}$ and H. Kachkachi$^{2}$}

\affiliation{${}^{1}$ \mbox{Physics Department, Lehman College, City University
of New York,} \\
 \mbox{250 Bedford Park Boulevard West, Bronx, New York 10468-1589,
USA}\\
 ${}^{2}$ \mbox{Laboratoire de Math{\'e}matiques, Physique et
Syst{\`e}mes, Universit{\'e} de Perpignan,} \\
 \mbox{52 Avenue de Paul Alduy, 66860 Perpignan Cedex, France}}

\date{\today}
\begin{abstract}
By numerically solving the equations of motion for atomic spins we
show that internal spin-wave processes in large enough magnetic particles,
initially in unstable states, lead to complete magnetization reversal
and thermalization. The particle's magnetization strongly decreases
in the middle of reversal and then recovers. We identify two main
scenarios, exponential and linear spin-wave instabilities. For the
latter, the longitudinal and transverse relaxation rates have been
obtained analytically. Orientation dependence of these rates leads
to a nonexponential relaxation of the particle's magnetization at
long times.
\end{abstract}

\pacs{75.50.Tt, 75.75.+a, 75.30.Ds, 76.20.+q}

\maketitle
%\pacs{75.50.Tt, 75.75.+a, 75.30.Ds, 76.20.+q}

\section{Introduction}

From both fundamental and application viewpoints, the switching
time required for the magnetization reversal has become of utmost
significance today. The tremendous increase in storage density and
read-write speed in magnetic storage media is reaching its limits
and the need for further speedup of the magnetization dynamics is
one of the main issues. Actual read-write processes operate on the
order of nanoseconds and newly developed methods such as
precessional switching \cite{acremannetal00science} have
demonstrated the possibility to reach sub-nanosecond time scales.
Recent pump-probe experiments have also shown a fast decay of the
magneto-optical signal occurring on the sub-picosecond time scale
using time-resolved techniques such as, to cite a few, the
magneto-optical Kerr effect (MOKE), \cite{beaurepaireetal96prl}
time-resolved second harmonic generation (SHG),
\cite{hohlfeldetal97prl} and pulsed inductive microwave
magnetometer (PIMM). \cite{schgerkossil05jap}

These techniques also make it possible to prepare the system in a
non-equilibrium state with the magnetization vector pointing in a
freely chosen arbitrary direction. In particular, unlike linear
FMR experiments, large-angle motion of the magnetization can be
studied. This is relevant to technological applications since
magnetic recording heads are expected to change magnetization
directions over large angles at ever increasing frequencies.
Another advantage of these techniques is that they are sensitive
to spin-wave and damping phenomena.
\cite{vankampenetal02prl,koopmansetal05prl,lounalkle05prbrc}

In this paper we will study the relaxational switching of the
magnetization initially put in a state far from equilibrium with
the methods mentioned above.

In general, relaxation is accompanied by a change of the energy of
the relaxing system, and the energy released or absorbed has to be
accommodated by a larger system serving as a heat bath. In particular,
magnetization reversal of (quasi-single-domain) magnetic particles
towards equilibrium reduce their Zeeman and/or anisotropy energy,
and the role of heat bath can be played by phonons and/or conduction
electrons, etc. Most of the existing theories consider magnetic particles
as \emph{single macroscopic magnetic moments} whose dynamics is described
by the Landau-Lifshitz equation with damping\cite{lanlif35} and Langevin
noise field (see, e.g., Refs. \onlinecite{bro63pr,gar97prb,garfes04prb}).

As was pointed out by Suhl \cite{suhl98ieee}\ and demonstrated by
Safonov and Bertram, \cite{safber01prb,safonov04jap} internal spin-wave
(SW) modes in the particle can serve as a heat bath and thus be responsible
for the particle's relaxation far from equilibrium and thereby for
magnetization switching. As a rule, this internal relaxation process
should be much faster than those processes that operate via nonmagnetic
degrees of freedom, with important implications in engineering of
magnetic elements in electronics. Progress in computing allowed to
simulate magnetic particle's dynamics as that of a system of \emph{many
interacting spins}, \cite{safber01prb,safonov04jap} considered classically.
Simulations show that relaxation via internal SW indeed occurs.

From the theoretical viewpoint, SW dynamics in a magnetic particle
whose global magnetization strongly depends on time in the course
of reversal is a new and challenging subject. Dynamics involving \emph{moderate}
deviations of the particle's magnetization from equilibrium, including
the Suhl SW instabilities \cite{suhl57jpcs} that may lead to a much
faster relaxation, \cite{suhl98ieee,dobvic03prl,lounalkle05prbrc}
can still be described with the help of a nonlinear SW theory built
around the ground state. To the contrast, large deviations require
redefinition of the spin-wave vacuum by considering SW dynamics in
the frame related to the instantaneous particle's global magnetization.
The vacuum corresponding to a collinear state with all particle's
spins oriented in an arbitrary direction is an example of a \emph{false
vacuum} that is unstable and decays toward the true vacuum. A spin-wave
vacuum of this kind was suggested for the thermodynamics of low-dimensional
magnetic systems in zero field that have a zero order parameter at
$T>0.$ \cite{hasnie93zpb} Its extension for nonzero field was done
in Ref.~\onlinecite{kacgar01epjb}. Recently, the initial stages
of SW dynamics including instabilities have been considered in the
global-magnetization frame in Ref.~\onlinecite{kas06prl}.

The false-vacuum initial condition can be created by biasing the system
with a magnetic field that leads to the disappearance of the metastable
energy minimum in which the magnetization is set. \cite{safber01prb}
Other methods are the precessional\cite{backetal99science} and current-induced
\cite{oezetal03prl} switching that allow to rotate the magnetization
into an arbitrary direction during the time that is shorter than the
relaxation time.

The aim of this paper, which is the extended version of the preceding
letter \cite{garkacrey08epl}, is two-fold:
\begin{enumerate}
\item We show by numerical simulations for atomic spins on the lattice at
initially $T=0$ that, for particles large enough, excitation of internal
spin waves leads to a full magnetization switching and relaxation
to a thermal state with the temperature $T$ defined from the energy
balance. This solves the puzzle of {}``incomplete relaxation'' observed
in Ref.~\onlinecite{safber01prb} for a particle of only 64 effective
spins. We show that magnetization switching via internal spin waves
is typically accompanied by a strong reduction of the particle's magnetization
$m$ that subsequently recovers to a value that is slightly less than
the initial value $m=1$ for collinear spins. This longitudinal relaxation
is similar to that described by the Landau-Lifshitz-Bloch (LLB) equation
(see, e.g., Refs.~\onlinecite{gar97prb,garfes04prb}), although
it has a different origin.
\item In addition to the \emph{exponential instability} mechanism that occurs
in the case of elliptic magnetization precession \cite{suhl57jpcs,suhl98ieee,safber01prb,dobvic03prl,kas06prl},
there occurs a \emph{linear instability} mechanism for the circular
magnetization precession. The latter may be driven by a random core
(volume) anisotropy or by a surface anisotropy that causes spin non-collinearities.
We analytically derive the rates describing the creation of SWs out
of the false vacuum, using the spin-wave theory in the frame of the
particle's global magnetization $\mathbf{m}$. Essential dependence
of these rates on the angle between $\mathbf{m}$ and the magnetic
field $\mathbf{H}$, which shapes the energy landscape, leads to a
slow non-exponential relaxation of $\mathbf{m}$ at large times. Furthermore,
depending on the direction of the applied field, the magnetization
switching may or may not occur.
\end{enumerate}
The remainder of the paper is organized as follows. Sec. \ref{Sec-Ham}
introduces the classical-spin model of a magnetic particle including
a bulk anisotropy and a random anisotropy. Here the existence of two
types of SW instabilities is demonstrated. Sec. \ref{Sec-Anal} contains
the formalism of spin-wave theory in the frame related to particle's
global magnetization. The details of the full analytical solution
in the case of random anisotropy are given in the Appendix. In Sec.
\ref{Sec-Num} the results of numerical atomistic simulations are
presented that show the two instability scenarios leading to the magnetization
reversal. Sec. \ref{Sec-Disc} contains our discussion.

\section{The Hamiltonian and spin-wave instabilities}

\label{Sec-Ham}

\subsection{The Hamiltonian}

We consider the classical Hamiltonian ($|\mathbf{s}_{i}|=1$) on the
lattice \begin{equation}
\mathcal{H}=\sum_{i}\mathcal{H}_{Ai}-\mathbf{h}\cdot\sum_{i}\mathbf{s}_{i}-\frac{1}{2}\sum_{ij}J_{ij}\mathbf{s}_{i}\cdot\mathbf{s}_{j},\label{Ham}\end{equation}
 where $\mathbf{h}=\mu_{0}\mathbf{H,}$ $\mu_{0}$ is the magnetic
moment associated with the spin, $\mathbf{H}$ is the magnetic field,
$J_{ij}$ is the exchange interaction, and $\mathcal{H}_{Ai}$ is
the crystal-field energy at site $i,$ a function of $\mathbf{s}_{i}$
satisfying the symmetry of the problem. In applications below, we
will consider the\ bulk uniaxial anisotropy with easy axis $\mathbf{e}_{z}$
\begin{equation}
\mathcal{H}_{Ai}=-D\left(\mathbf{e}_{z}\cdot\mathbf{s}_{i}\right)^{2},\qquad D>0\label{HamAUniax}\end{equation}
 and the random anisotropy \begin{equation}
\mathcal{H}_{Ai}=-\sum_{\alpha\beta}g_{i,\alpha\beta}s_{i\alpha}s_{i\beta}\label{HamARand}\end{equation}
 with \begin{equation}
g_{i,\alpha\beta}=D_{R}\left(u_{i\alpha}u_{i\beta}-\frac{1}{3}\delta_{\alpha\beta}\right),\qquad D_{R}>0\label{gRADef}\end{equation}
 $\mathbf{u}_{i}$ being a unit vector assuming random directions.\cite{harplizuc73prl,chudnovsky95WS}
One can also add a surface anisotropy and dipole-dipole interaction.
Throughout the paper we do not include any coupling to the environment.

The particle's magnetization is defined as \begin{equation}
\mathbf{m}=\frac{1}{\mathcal{N}}\sum_{i}\mathbf{s}_{i},\label{mDef}\end{equation}
 where $\mathcal{N}$ is the total number of spins.

Atomic spins obey the Larmor equation \begin{equation}
\mathbf{\dot{s}}_{i}=\left[\mathbf{s}_{i}\times\mathbf{\Omega}_{i}\right],\qquad\hbar\mathbf{\Omega}_{i}=-\partial\mathcal{H/}\partial\mathbf{s}_{i}.\label{LLEqi}\end{equation}

\subsection{Spin-wave instabilities}

We now study two models, one with uniaxial anisotropy with the same
easy axis for all spins and no random anisotropy ($D_{R}=0$), and
the other with random anisotropy and no uniaxial anisotropy ($D=0$).
We will show that in the case of uniaxial anisotropy the SW instabilities
are exponential while in the case of random anisotropy these instabilities
are linear. The calculations are both analytical and numerical in
the latter case and only numerical in the former.

\subsubsection{\label{sub:UA+EI}Uniaxial anisotropy and exponential instabilities}

The first model we study is that of uniaxial anisotropy with the common
easy axis in the $z$ direction, noncollinear with the applied field
$\mathbf{h}$. In this case, the spin precession is non-circular.
The same effect may also be caused by a \emph{biaxial} anisotropy.
We choose the initial state $\mathbf{s}_{i}=\mathbf{e}_{x}$ and $\mathbf{h=}h\mathbf{e}_{x}.$
Linearization around this state yields the SW spectrum \begin{equation}
\varepsilon_{\mathbf{k}}=\sqrt{\left(h+J_{0}-J_{\mathbf{k}}\right)\left(h-2D+J_{0}-J_{\mathbf{k}}\right)},\label{omegakUniaxperpH}\end{equation}
 where $J_{\mathbf{k}}$ is the Fourier coefficient of $J_{ij}$ which,
for a particle with simple cubic (sc) structure, reads \begin{equation}
J_{\mathbf{k}}=2J\sum_{\alpha}\cos(ak_{\alpha}),\label{Jksc}\end{equation}
 where $a$ is the lattice spacing. In the long-wave-length limit
$J_{0}-J_{\mathbf{k}}\cong J\left(ak\right)^{2}.$

Let us now consider the properties of Eq. (\ref{omegakUniaxperpH})
in different regions of the magnetic field $h$.
\begin{itemize}
\item For $h>2D$ the state $\mathbf{s}_{i}=\mathbf{e}_{x}$ is the energy
minimum and thus $\varepsilon_{\mathbf{k}}$ is real. This means that
there is no SW instabilities and the initially excited SW will maintain
their amplitudes.
\item In the interval $0<h<2D$ the state $\mathbf{s}_{i}=\mathbf{e}_{x}$
is a saddle point, so that $\mathbf{m}$ can rotate away from this
state, similarly to the case studied in Ref.~\onlinecite{safber01prb}.
SW modes in the interval $0<$ $J_{0}-J_{\mathbf{k}}<2D-h$ are unstable
since $\varepsilon_{\mathbf{k}}$ becomes imaginary. \cite{kas06prl}
This leads to the exponential increase of the deviations from the
initially nearly collinear state. The highest instability increment
is realized in the middle of the $\mathbf{k}$-interval of instability,
i.e., $J_{0}-J_{\mathbf{k}}=D-h/2.$ This means that exponentially
growing spin waves with this nonzero value of $\mathbf{k}$ dominate
in the instability process. As a result, the magnetization length
$m\equiv|\mathbf{m}|$ decreases upon rotation out of the saddle point,
as was observed in early simulations.\cite{safber01prb}
\item For $h<0$ the state $\mathbf{s}_{i}=\mathbf{e}_{x}$ is the energy
maximum and thus $\mathbf{m}$ performs small-amplitude precession
around $\mathbf{e}_{x}$ in the absence of SW processes. Indeed, according
to Eq.\ (\ref{omegakUniaxperpH}), the mode $\mathbf{k=0}$ is stable
as $\varepsilon_{\mathbf{0}}$ is real. On the other hand, $\mathbf{k\neq0}$
modes in the interval\\
\begin{equation}
-h<J_{0}-J_{\mathbf{k}}<2D-h\label{eq:InstabilityRange}\end{equation}
\\
are unstable. In this case the only way of reversal is via excitation
of internal spin waves with $\mathbf{k\neq0}$ that strongly reduce
the magnetization magnitude $m$. This means that the uniform-precession
mode decays in favor of non-uniform spin-wave modes.
\end{itemize}

\subsubsection{\label{sub:RA+LI}Random anisotropy and linear instabilities}

The second situation we consider here is with $D=0$ and non-zero
random anisotropy of Eq.\ (\ref{HamARand}). The latter does not
break the global particle's magnetic isotropy, and its only role is
to provide strength for SW conversion processes that lead to non-conservation
of $\mathbf{m}\mathbf{\cdot h}$ and thus to reversal. We choose $\mathbf{h=}h\mathbf{e}_{z}$
with $h>0$. Here the precession of $\mathbf{m}$ is circular and
there are no exponential SW instabilities. Instead, spin waves may
be generated out of a false vacuum by linear transformation processes.
In particular, for $\mathbf{m}$ antiparallel to $\mathbf{h}$, spin
waves in the particle have a negative gap $-h$. Thus a SW with $\mathbf{k\neq0}$
can be created out of the false vacuum if its energy is zero: \begin{equation}
\varepsilon_{\mathbf{k}}=-h+J_{0}-J_{\mathbf{k}}=0.\label{ResCondSW}\end{equation}
 In Sec. \ref{Sec-Anal} and the Appendix this process will be considered
in detail for any angle between $\mathbf{m}$ and $\mathbf{h.}$ The
amplitudes of unstable SW and the deviation of $m$ from saturation
increase linearly with time at small times.

\subsubsection{Size effects}

In the analysis of both exponential and linear SW instabilities one
has to take into account the fact that because of the finite size
of magnetic particles their internal SW modes are discrete. In particular,
for a box-shaped particle with free boundary conditions (fbc) there
are standing spin waves with wave vectors \cite{kacgar01epjb} \begin{equation}
k_{\alpha}=\frac{\pi n_{\alpha}}{aN_{\alpha}},\quad n_{\alpha}=0,1,\ldots,N_{\alpha}-1,\quad\alpha=x,y,z,\label{kfbc}\end{equation}
 where $a$ is the lattice spacing and $N_{x}N_{y}N_{z}=\mathcal{N}$
is the total number of spins. For most of other shapes, SW modes in
magnetic particles have to be found numerically and they are labeled
by discrete wave numbers rather than by the wave vector $\mathbf{k.}$
In the sequel, in the analytical calculations, we will consider only
the box-shaped particles for simplicity.

One can see that for the linear instability process, Eq. (\ref{ResCondSW})
may be satisfied for a particular SW mode. For particles small enough,
the lowest value of $J_{0}-J_{\mathbf{k}}$ for $\mathbf{k\neq0}$
{[}i.e., for $n_{\alpha}=1$ in Eq.\ (\ref{kfbc})] exceeds $h$
and thus SWs cannot be created. This yields the absolute stability
criterion for the particle's linear size $L$ \begin{equation}
L<L^{\ast}=aN^{\ast}=\pi a\sqrt{J/h},\label{LinearStabCrit}\end{equation}
 assuming a cubic shape. If there is only one mode that exactly or
approximately satisfies Eq.\ (\ref{ResCondSW}) and this mode does
not decay into second-generation SW, there are harmonic oscillations
between the false-vacuum state and the state with the resonant SW
mode: $1-m\sim\cos(\Omega t),$ where $\Omega$ depends on the strength
of the SW conversion processes, i.e., on the random anisotropy or
other interactions creating spin non-collinearity. These oscillations
are similar to the probability oscillations between two resonant states
in quantum mechanics. If the unstable spin-wave mode can be converted
into a second-generation spin-wave mode that also has its energy close
to zero, the relaxation process becomes a two-step process, and the
temporal behavior of $\mathbf{m}$ becomes more complicated. For small
enough particles, but with $L\gtrsim L^{\ast},$ there are not enough
resonant SW modes, so that the relaxation process gets stuck in what
can be called {}``spin-wave bottleneck'' and the relaxation is incomplete,
as was observed in Ref \onlinecite{safber01prb}. In large particles,
$L\gg L^{\ast},$ SW modes become quasi-continuous, the SW bottleneck
disappears, and a cascade of SW processes leads to a nearly full magnetization
reversal, as is demonstrated below.

Analysis of the exponential instability in magnetic particles of finite
size goes along similar lines. The particle is \emph{stable} if the
smallest value of $J_{0}-J_{\mathbf{k}}$ with $\mathbf{k\neq0}$
(i.e., $n_{\alpha}=1)$ exceeds the right boundary of the instability
interval, i.e., $2D-h$, see Eq. (\ref{eq:InstabilityRange}). This
leads to the stability criterion \begin{equation}
L<L^{\ast}=aN^{\ast}=\pi a\sqrt{\frac{J}{2D+|h|}}\label{SingleDomainCrit}\end{equation}
 for the particle's size $L=\max(L_{x},L_{y},L_{z}).$ Eq.~(\ref{SingleDomainCrit})
is similar to the single-domain criterion since its right-hand side
is the domain-wall width, if $h=0$. If $L\gtrsim L^{\ast}$ and there
is only one SW mode inside the instability interval that does not
convert into second-generation spin waves, its amplitude and thus
the magnetization length $m$ depends periodically on time. This dependence
is not sinusoidal, as that for the linear instability, see comment
below Eq. (\ref{LinearStabCrit}). Instead, the SW amplitude initially
exponentially increases but then this evolution becomes inverted due
to nonlinear effects at large amplitudes, and the SW amplitude reversibly
returns to the starting small value. After that the process repeats
periodically. If second-generation unstable spin waves are created,
the time evolution of $\mathbf{m}$ becomes more complicated and $m$
does not return to saturation. Still, for $L\gtrsim L^{\ast}$ the
magnetization reversal is incomplete, as observed in Ref. \onlinecite{safber01prb}.
Only for large enough particles, $L\gg L^{\ast},$ does relaxation
via internal spin waves lead to a complete magnetization reversal.

Finally, we address the question as to how large the particle must
be so that its spectrum of spin waves becomes effectively continuous.
In the case of the exponential instability the criterion is that there
must be several modes within the instability interval whose length
(in energy) is proportional to $D,$ as we have seen above. Thus the
criterion of the quasi-continuous spectrum becomes \begin{equation}
\Delta\varepsilon\lesssim D,\label{CritQuasiCont}\end{equation}
 where $\Delta\varepsilon$ is the average distance between the SW
modes. In the case of random anisotropy, the spin-wave deviations,
associated with a particular mode, satisfy linear differential equations
with a source that is proportional to the random anisotropy constant
$D_{R},$ see e.g., Eq. (\ref{zetakEq}). The response to the off-resonance
source is of order $D_{R}/\Delta\varepsilon$ that becomes large for
$\Delta\varepsilon\lesssim D_{R},$ formally coinciding with Eq. (\ref{CritQuasiCont}).
If this condition is fulfilled, spin waves are effectively generated
for all values of the field $h$ without the necessity to satisfy
Eq. (\ref{ResCondSW}). Next, $\Delta\varepsilon$ can be expressed
via the density of SW states, defined by Eq. (\ref{rhoDef}), as \begin{equation}
\Delta\varepsilon=\frac{\hbar}{\mathcal{N}\rho(\omega)}\end{equation}
 in the vicinity of the energy $\varepsilon=\hbar\omega.$ Thus Eq.
(\ref{CritQuasiCont}) can be rewritten in terms of the particle's
size as \begin{equation}
\mathcal{N}\gtrsim\frac{\hbar}{\rho(\omega)D}.\label{NContCrit}\end{equation}
 Adopting the rough estimation $\hbar/\rho(\omega)\sim J$ over the
whole range of SW frequencies, one obtains the criterion of the quasi-continuous
spectrum in the form \begin{equation}
\mathcal{N}\gtrsim J/D\label{NContCritRough}\end{equation}
 and the same with $D\Rightarrow D_{R}$ for random anisotropy.

\section{Analytical theory of spin-wave instabilities in the particle frame}

\label{Sec-Anal}

In this section, we present our general formalism of SW theory in
the frame related with the particle's global magnetization for an
arbitrary direction of the applied field. Then, we study the two anisotropy
models of sections \ref{sub:UA+EI} and \ref{sub:RA+LI} and the ensuing
SW instabilities.

The microscopic effective field $\hbar\mathbf{\Omega}_{i}$ in Eq.~(\ref{LLEqi})
can be written in the form \begin{equation}
\hbar\mathbf{\Omega}_{i}=\mathbf{h}-\frac{\partial\mathcal{H}_{Ai}}{\partial\mathbf{s}_{i}}+2\overleftrightarrow{\mathbf{g}}_{i}\mathbf{s}_{i}+\sum_{j}J_{ij}\mathbf{s}_{j},\label{Omegai}\end{equation}
 where $\mathcal{H}_{Ai}$ contains only non-random anisotropy, whereas
the random anisotropy is singled out. In particular, for $\mathcal{H}_{Ai}$
given by Eq. (\ref{HamAUniax}) one has $-\partial\mathcal{H}_{Ai}$/$\partial\mathbf{s}_{i}$=$\overleftrightarrow{\mathbf{D}}\mathbf{s}_{i}\equiv$
$\overleftrightarrow{\mathbf{D}}\cdot\mathbf{s}_{i}$ and the components
of the tensor $\overleftrightarrow{\mathbf{D}}$ read $(\overleftrightarrow{\mathbf{D}})_{\alpha\beta}=D\delta_{\alpha z}\delta_{\beta z}.$
Similarly, the components of the random-anisotropy tensor are given
by $(\overleftrightarrow{\mathbf{g}_{i}})_{\alpha\beta}=g_{i,\alpha\beta}$
and Eq. (\ref{gRADef}). One can represent $\mathbf{s}_{i}$ in the
form \begin{equation}
\mathbf{s}_{i}=\mathbf{m}+\mathbf{\psi}_{i},\qquad\sum_{i}\mathbf{\psi}_{i}=0,\label{psiDef}\end{equation}
 where $\mathbf{m}$ is the average spin defined by Eq.\ (\ref{mDef})
and $\mathbf{\psi}_{i}$ contains the Fourier components with $\mathbf{k\neq0}$
and describes spin waves in the particle. Whereas in the standard
SW theory $\mathbf{m}$ is a constant corresponding to the ground-state
orientation, here it is treated as a time-dependent variable. Since
the atomic spins are subject to the chiral constraint $\mathbf{s}_{i}^{2}=1,$
one can use \cite{kas06prl} $\mathbf{m=n}\sqrt{1-\psi_{i}^{2}}$
with $\mathbf{n}\cdot\mathbf{\psi}_{i}=0,$ where $\mathbf{n}$ is
a unit vector. Although this reduces to two the number of the $\mathbf{\psi}_{i}$
components to deal with, the formalism becomes much more cumbersome,
the final results, however, are not affected. Thus we decided not
to use the chiral constraint explicitly in our presentation. Of course,
properly written equations must satisfy this constraint that can be
used to check them.

The equation of motion for $\mathbf{m}$ following from Eq.\ (\ref{LLEqi})
has the form \begin{equation}
\hbar\mathbf{\dot{m}}=\left[\mathbf{m\times h}_{\mathrm{eff}}\right]+\mathbf{R.}\label{mEq0}\end{equation}
 Here \begin{equation}
\mathbf{h}_{\mathrm{eff}}\equiv\mathbf{h}-\left.\frac{\partial\mathcal{H}_{Ai}}{\partial\mathbf{s}_{i}}\right|_{\mathbf{s}_{i}\Rightarrow\mathbf{m}}\label{heffDef}\end{equation}
 is the effective field acting on the particle as a whole and does
not contain the random anisotropy and exchange coupling. The term
$\mathbf{R}$ couples the dynamics of $\mathbf{m}$ to that of spin
waves described by $\mathbf{\psi}_{i}$ and it is responsible for
the relaxation of $\mathbf{m.}$ Calculation yields \begin{eqnarray}
\mathbf{R} & = & \frac{1}{\mathcal{N}}\sum_{i}\left\{ \left[\mathbf{m\times}2\overleftrightarrow{\mathbf{g}}_{i}\mathbf{\psi}_{i}\right]\right.\notag\\
 &  & \quad+\left.\left[\mathbf{\psi}_{i}\times\left(\overleftrightarrow{\mathbf{F}}\mathbf{\psi}_{i}+2\overleftrightarrow{\mathbf{g}}_{i}\left(\mathbf{m+\psi}_{i}\right)\right)\right]\right\} .\label{RDef}\end{eqnarray}
 Here $\sum_{i}\left[\mathbf{m\times}2\overleftrightarrow{\mathbf{g}}_{i}\mathbf{m}\right]=0$
as an average of the random anisotropy, $\sum_{ij}J_{ij}\left[\mathbf{\psi}_{i}\times\mathbf{\psi}_{j}\right]=0$
by symmetry, while some other terms vanish by virtue of Eq. (\ref{psiDef}).
The tensor $\overleftrightarrow{\mathbf{F}}$ is given by \begin{equation}
\left(\overleftrightarrow{\mathbf{F}}\right)_{\alpha\beta}=-\left.\frac{\partial^{2}\mathcal{H}_{Ai}}{\partial s_{i\alpha}\partial s_{i\beta}}\right|_{\mathbf{s}_{i}\Rightarrow\mathbf{m}}\label{FtensDef}\end{equation}
 For $\mathcal{H}_{Ai}$ given by Eq. (\ref{HamAUniax}) one has $\left(\overleftrightarrow{\mathbf{F}}\right)_{\alpha\beta}=2(\overleftrightarrow{\mathbf{D}})_{\alpha\beta}=2D\delta_{\alpha z}\delta_{\beta z}.$

In turn, the equation of motion for $\mathbf{\psi}_{i}$ can be obtained
as \begin{equation}
\mathbf{\dot{\psi}}_{i}=\mathbf{\dot{s}}_{i}\mathbf{-\dot{m}=}\left[\mathbf{s}_{i}\times\mathbf{\Omega}_{i}\right]\mathbf{-\dot{m}.}\label{psiibasicEq}\end{equation}
 Working out the various terms yields \begin{equation}
\hbar\mathbf{\dot{\psi}}_{i}=\left[\mathbf{m\times}2\overleftrightarrow{\mathbf{g}}_{i}\mathbf{m}\right]+\mathbf{A}_{i}^{(1)}+\mathbf{A}_{i}^{(2)},\label{psiiEq}\end{equation}
 where $\mathbf{A}_{i}^{(1)}$ contains terms linear in $\mathbf{\psi}_{i}$
and not containing $\overleftrightarrow{\mathbf{g}}_{j},$ while $\mathbf{A}_{i}^{(2)}$
contains the terms of order $\mathbf{\psi}^{2}$ and $\mathbf{\psi g.}$
In the sequel we will keep only $\mathbf{A}_{i}^{(1)}$ that is responsible
for the generation of spin waves out of a false vacuum, whereas $\mathbf{A}_{i}^{(2)}$
responsible for nonlinear spin-wave processes will be dropped. One
has \begin{eqnarray}
\mathbf{A}_{i}^{(1)} & = & \left[\mathbf{\psi}_{i}\times\left(\mathbf{h}_{\mathrm{eff}}+J_{0}\mathbf{m}\right)\right]\notag\\
 &  & \qquad+\left[\mathbf{m\times}\left(\overleftrightarrow{\mathbf{F}}\mathbf{\psi}_{i}+\sum_{j}J_{ij}\mathbf{\psi}_{j}\right)\right].\label{A1iDef}\end{eqnarray}

The first term in Eq.\ (\ref{psiiEq}) that causes non-collinearity
of spins, is responsible for the linear SW instability. The same effect
is produced by surface anisotropy and dipole-dipole interaction. This
source term in the linear equation for $\mathbf{\psi}_{i}$ is due
to the lack of translational invariance of the random anisotropy.
On the contrary, the terms due to the bulk anisotropy $\mathcal{H}_{Ai}$
and magnetic field $\mathbf{h}$\ are invariant by translation and
thus enter the equation of motion for the particle's global magnetization
$\mathbf{m}$ \ (\ref{mEq0}) rather than the equation for $\mathbf{\psi}_{i}.$

In Eq. (\ref{mEq0}) it is convenient to project the relaxation term
$\mathbf{R}$ onto $\mathbf{m}$ and the perpendicular directions.
For this we introduce orthogonal unit vectors \begin{eqnarray}
\mathbf{n} & = & \frac{\mathbf{m}}{m}\notag\\
\mathbf{e}_{1} & = & \frac{\left[\mathbf{n\times h}_{\mathrm{eff}}\right]}{\left|\left[\mathbf{n\times h}_{\mathrm{eff}}\right]\right|}=\frac{\left[\mathbf{n\times h}_{\mathrm{eff}}\right]}{h_{\mathrm{eff}}\sqrt{1-x^{2}}}\notag\\
\mathbf{e}_{2} & = & \left[\mathbf{n}\times\mathbf{e}_{1}\right]=\frac{\left[\mathbf{n}\times\left[\mathbf{n}\times\mathbf{h}_{\mathrm{eff}}\right]\right]}{\left|\left[\mathbf{n}\times\mathbf{h}_{\mathrm{eff}}\right]\right|},\label{RotFrameDef}\end{eqnarray}
 where \begin{equation}
x\equiv\frac{\mathbf{n\cdot h}_{\mathrm{eff}}}{h_{\mathrm{eff}}}=\frac{\mathbf{m\cdot h}_{\mathrm{eff}}}{mh_{\mathrm{eff}}}.\label{xDef}\end{equation}
 $\mathbf{R}$ then has the form \begin{equation}
\mathbf{R=}R_{\Vert}\mathbf{n}+R_{1}\mathbf{e}_{1}+R_{2}\mathbf{e}_{2},\label{xivian0e1e2}\end{equation}
 where $R_{\Vert}$ describes the longitudinal relaxation of the particle's
magnetization (change in the magnetization magnitude) while \begin{equation}
R_{2}\equiv R_{\bot}\label{R2isRperp}\end{equation}
 describes the transverse relaxation (rotational relaxation of the
magnetization vector), since $\mathbf{e}_{2}\cdot\mathbf{n}=0$. On
the contrary, $R_{1}$ does not describe any relaxation; it merely
describes a small modification of the particle's precession due to
excitation of spin waves, an effect that will be neglected here. Explicitly
Eq. (\ref{mEq0}) can now be rewritten as \begin{equation}
\hbar\mathbf{\dot{m}}=\left[\mathbf{m\times h}_{\mathrm{eff}}\right]+R_{\Vert}\frac{\mathbf{m}}{m}+R\mathbf{_{\bot}}\frac{\left[\mathbf{m}\times\left[\mathbf{m}\times\mathbf{h}_{\mathrm{eff}}\right]\right]}{m\left|\left[\mathbf{m}\times\mathbf{h}_{\mathrm{eff}}\right]\right|}\mathbf{,}\label{mEqexpl}\end{equation}
 where \begin{equation}
R_{\Vert}=\mathbf{n\cdot R,\qquad}R\mathbf{_{\bot}=e}_{2}\mathbf{\cdot R.}\label{RparperpDef}\end{equation}
 Eq. (\ref{mEqexpl}) resembles the Landau-Lifshitz-Bloch (LLB) equation
since it includes both the longitudinal and transverse relaxation
terms. Ignoring $\mathbf{A}_{i}^{(2)}$ in Eq.\ (\ref{psiiEq}),
one can solve the resulting linear equation for $\mathbf{\psi}_{i}$,
insert the solution into $\mathbf{R}$, and obtain the relaxation
terms in Eq.\ (\ref{mEqexpl}) from Eq. (\ref{RparperpDef}). Integrating
out $\mathbf{\psi}_{i}$ can be done in the particle's frame defined
by Eq. (\ref{RotFrameDef}). It is understood that Eq. (\ref{mEqexpl})
is only valid during the initial stage of the evolution out of the
completely or nearly collinear state, $|\mathbf{m|}\cong1\mathbf{,}$
when the deviations $\mathbf{\psi}_{i}$ due to spin waves are small
and can be considered perturbatively.

\subsection{Linear instabilities}

Eq. (\ref{psiiEq}) without the contribution $\mathbf{A}_{i}^{(2)}$
can be solved analytically in the case of a pure random anisotropy,
$\mathcal{H}_{Ai}=0,$ since then the time dependence of the frame
vectors is a simple precession around the external field $\mathbf{h}$.
The details can be found in the Appendix. For $D=0$ the dependence
$\mathbf{m}(t)$ {[}neglecting $\mathbf{R}$ in Eq. (\ref{mEq0})]
is a circular precession, and thus the calculations simplify and lead
to the equation of motion \begin{equation}
\mathbf{\dot{m}}=\frac{1}{\hbar}\left[\mathbf{m\times h}\right]-m^{3/2}\Gamma_{\Vert}(x)\frac{\mathbf{m}}{m}-m^{3/2}\Gamma_{\bot}(x)\frac{\left[\mathbf{m\times}\left[\mathbf{m\times h}\right]\right]}{m^{2}h}.\label{LLBSWEq}\end{equation}
 Here the transverse and longitudinal relaxation rates depend on the
orientation of the particle's magnetization vector $x\equiv\left(\mathbf{m\cdot h}\right)/\left(mh\right)$:
\begin{eqnarray}
\Gamma_{\Vert}(x) & = & \frac{2}{15\pi\hbar}\frac{D_{R}^{2}}{J}\sqrt{\frac{h}{J}}\Phi_{\Vert}(x)\equiv\Gamma_{\Vert0}\Phi_{\Vert}(x)\label{GammaPar}\\
\Gamma_{\bot}(x) & = &
\frac{1}{5\pi\hbar}\frac{D_{R}^{2}}{J}\sqrt{\frac{h}{J}}\Phi_{\bot}(x)\equiv\Gamma_{\bot0}\Phi_{\bot}(x)\label{GammaPerp}\end{eqnarray}
 with \begin{eqnarray}
\Phi_{\Vert}(x) & = & \frac{\left(1-x\right)^{2}}{4}\left[\left(1+2x\right)^{2}+\sqrt{2}\left(1-x^{2}\right)\right]\label{PhiPar}\\
\Phi_{\bot}(x) & = & \frac{1-x}{6}\left[\left(1+2x\right)^{2}+\sqrt{2}(1-x)(2+x)\right].\label{PhiPerp}\end{eqnarray}
 The applicability of our method requires $\Gamma_{\bot},\Gamma_{\Vert}\ll\omega_{H}=h/\hbar$,
i.e., the relaxation of the magnetization (as well as the rate of
SW production) is much slower than the magnetization precession considered
as unperturbed in the first approximation above.

We note that Eq.\ (\ref{LLBSWEq}) is similar to the LLB equation.
\cite{gar97prb,garfes04prb} However, it is valid, in general, only
for short times, when the number of excited spin waves is still small
and $m\cong1,$ so that neglecting $\mathbf{A}_{i}^{(2)}$ in Eq.\ (\ref{psiiEq})
is justified. Correspondingly, $m$ in Eq.\ (\ref{LLBSWEq}) may
be replaced by 1. Of more consequence, however, is to introduce the
magnetization direction $\mathbf{n}$ and write $\mathbf{m=n}m.$
Then Eq. (\ref{LLBSWEq}) can be split into the two following equations
\begin{equation}
\mathbf{\dot{n}}=\frac{1}{\hbar}\left[\mathbf{n\times h}\right]-\Gamma_{\bot}(x)\frac{\left[\mathbf{n\times}\left[\mathbf{n\times h}\right]\right]}{h},\label{LLBn}\end{equation}
 where $x\equiv\mathbf{n\cdot h/}h$, and \begin{equation}
\dot{m}=-\Gamma_{\Vert}(x).\label{LLBm}\end{equation}

The small-$t$ behavior \begin{equation}
m_{z}(t)=1-\Gamma_{\Vert}(-1)t\label{mztinitial}\end{equation}
 following from Eq.\ (\ref{LLBm}) and shown in Fig.\ \ref{fig:Random-box}a.
It agrees well with the numerical result with a small discrepancy
stemming from $h/J$ not being small enough to use the analytical
expression for the density of states given by Eq. (\ref{rhoeps}).
Equation.\ (\ref{LLBm}) does not describe the increase of $m$ after
switching and its recovery to $m\cong1$ that is seen in Fig. \ref{fig:Random-box}a.

A striking feature of our result is that both $\Gamma_{\bot}(x)$
and $\Gamma_{\Vert}(x)$ vanish for $\mathbf{m}\Vert\mathbf{h}$ (i.e.,
for $x=1),$ while they reach their maxima at $x=-1$. Thus, initially
fast relaxation slows down when the particle approaches equilibrium,
see Fig.\ \ref{fig:Random-box}. Indeed, Eq. (\ref{LLBn}) can be
rewritten as \begin{equation}
\dot{x}=\Gamma_{\bot}(x)(1-x^{2}).\label{xdotEq}\end{equation}
 In terms of the angular deviation from equilibrium we have \begin{equation}
y\equiv1-\mathbf{n\cdot h/}h\equiv1-x\ll1\label{yDef}\end{equation}
 upon which Eq. (\ref{xdotEq}) simplifies into the equation $\dot{y}=-3\Gamma_{\bot0}y^{2}$
and whose solution reads \begin{equation}
y(t)=\frac{y(0)}{1+3y(0)\Gamma_{\bot0}t}.\label{ySol}\end{equation}

It is seen that the long-time asymptote of this solution does not
depend on the initial condition $y(0)$ and is given by $y(t)=1/\left(3\Gamma_{\bot0}t\right).$
The full magnetization vector $\mathbf{m=n}m$ includes $m$ that
follows from Eq. (\ref{LLBm}). The latter becomes $\dot{m}=-(9/4)\Gamma_{\Vert0}y^{2}=-(3/2)\Gamma_{\bot0}y^{2}=\dot{y}/2$.
This yields \begin{equation}
1-m(t)=\frac{1}{2}\left[y(0)-y(t)\right].\label{mtRes}\end{equation}

Thus, we see that the deviation $1-$ $m(t)$ remains finite and small
for $t\rightarrow\infty$ since $y(t)\rightarrow1/\left(3\Gamma_{\bot0}t\right)$,
that is the consequence of the relaxation slowing down as $\mathbf{m}$
approaches equilibrium. The change of $\mathbf{m}$ due to its rotation
and due to the change in its magnitude are comparable with each other.
In addition, one has to remember that the equation of motion for $\mathbf{m}$
was obtained for the initial stage of relaxation only and, in particular,
Eq. (\ref{LLBm}) does not describe thermalization. Still the evidence
provided by these analytical calculations of the non-exponential relaxation
via internal spin waves in magnetic particles is quite convincing.

\subsection{Exponential instabilities}

In the case of the purely bulk anisotropy (no random anisotropy and
other interactions causing noncollinearity of spins), Eq. (\ref{psiiEq})
with $\mathbf{A}_{i}^{(2)}\Rightarrow0$ is a system of uniform linear
equations. In general, $\mathbf{m}$ has a nontrivial quasi-periodic
time dependence defined by the bulk anisotropy, magnetic field, and
the energy associated with $\mathbf{m}$ that is conserved in the
absence of the internal spin waves. Excitation of the latter reduces
the energy associated with $\mathbf{m}$, so that the total energy
is conserved. Mathematically spin-wave instabilities in this case
are exponential divergences of the solution of a system of linear
differential equations with periodic coefficients. A well-known example
of such instabilities is parametric resonance. In all cases instability
requires elliptic or more complicated precession of $\mathbf{m.}$
Simple circular precession around the field $\mathbf{h}$ does not
lead to exponential instabilities.

From the physical point of view, exponential SW instabilities are
similar to the Suhl exponential instabilities. The difference resides
mainly in the mathematical description. The Suhl formalism considers
spin waves above the true ground state, so that it cannot be applied
in situations of large deviations of $\mathbf{m}$ from the energy
minimum. The advantage of this method is that unstable spin waves
are described by a linear differential equation with constant coefficients
that can easily be solved.

In contrast, the present method using the frame related to the particle's
global magnetization can be used for any deviations of $\mathbf{m}$
from the ground state. The price to pay is to deal with differential
equations with time-dependent coefficients that are difficult to solve
analytically (see also discussion in Ref. \onlinecite{kas06prl}).
However, if $\mathbf{m}$ is initially oriented towards an energy
maximum or a saddle point so that it does not evolve in time in the
absence of spin waves, one obtains differential equations for the
deviations $\mathbf{\psi}_{i}$ with constant coefficients that can
easily be solved showing exponential instabilities. In fact, in this
case one can just linearize Eq. (\ref{LLEqi}) near a given direction
of $\mathbf{m,}$ instead of going through the formalism of this section.
Examples are considered below Eq. (\ref{omegakUniaxperpH}).

\section{Numerical results}

\label{Sec-Num}

\begin{figure}
\includegraphics[angle=-90,width=8.5cm]{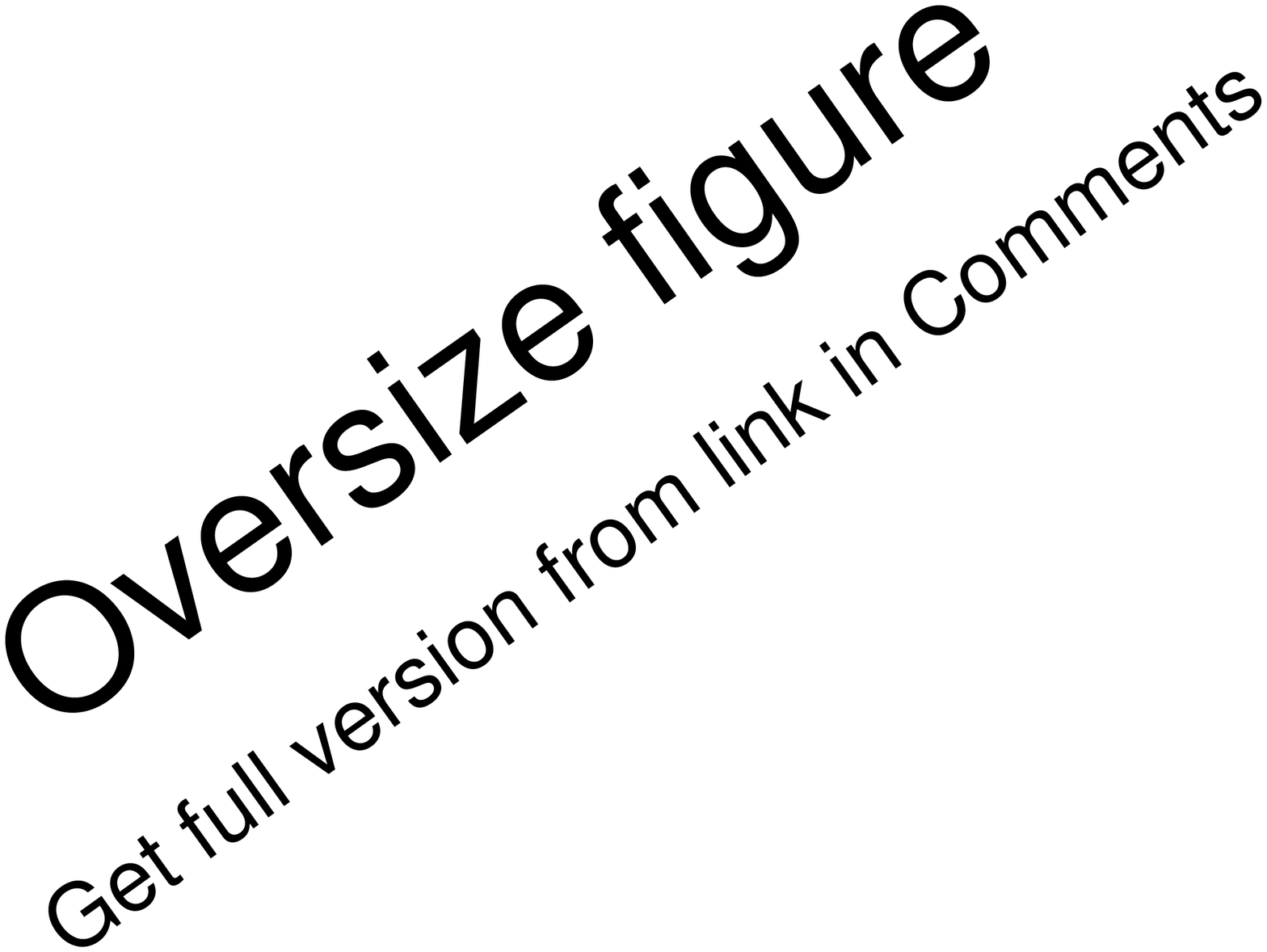}
\includegraphics[angle=-90,width=8.5cm]{Oversize_figure.eps}

\caption{\label{fig:Uniaxial-sphere}Magnetization switching out of the maximal-energy
state via exponential spin-wave instability in a spherical particle
with uniaxial anisotropy and transverse field. Switching occurs via
longitudinal relaxation without rotation. (b) shows the short-time
magnification. The {}``Plateau'' at $t\lesssim2000$ describes the
initial exponential increase of $1-m$.}

\end{figure}

Efficient magnetization reversal via the SW instability requires a
quasi-continuous SW spectrum, i.e., the existence of many SW modes
in the instability interval. For this, a \emph{strong} inequality
opposite to Eq.~(\ref{SingleDomainCrit}) should be fulfilled, i.e.,
the particle's size and/or the anisotropy $D$ should be large enough.
To illustrate the process, we have solved Eq.\ (\ref{LLEqi}) for
magnetization switching using the C++-based package \emph{MagneticParticle}
(\copyright\ H. Kachkachi \& L. Reynaud).

For the model with uniaxial anisotropy with easy axis along $z$,
the calculations have been done for two different particle's shapes.
One is a sphere of radius $R=35$ lattice spacings, cut inside a $70\times70\times70$
cube resulting in 171712 atomic spins. The other particle is a parallelepiped
consisting of $28\times31\times34=29512$ atomic spins. The idea behind
choosing all different sizes is to avoid degeneracy in the energies
of SW modes {[}see Eq. (\ref{kfbc})] and ensure a smoother density
of states and thus facilitate SW conversion processes. In both cases
the initial state is collinear and a very small surface anisotropy
was added to create initial very small deviations from the collinearity
that then exponentially grow. The lattice structure is simple cubic
and boundary conditions are free (fbc). The magnetic field is $\mathbf{h}=h\mathbf{e}_{x}$,
with $h>2D$ that creates the energy maximum in the direction opposite
to $\mathbf{h.}$ The latter is the initial direction of the magnetization
in part of the simulations. As stressed above, a particle whose magnetization
is oriented toward the energy maximum cannot just rotate out of this
orientation (in the absence of a coupling to the environment) because
of the energy conservation.

\begin{figure}[t]
\begin{centering}
\includegraphics[angle=-90,width=8.5cm]{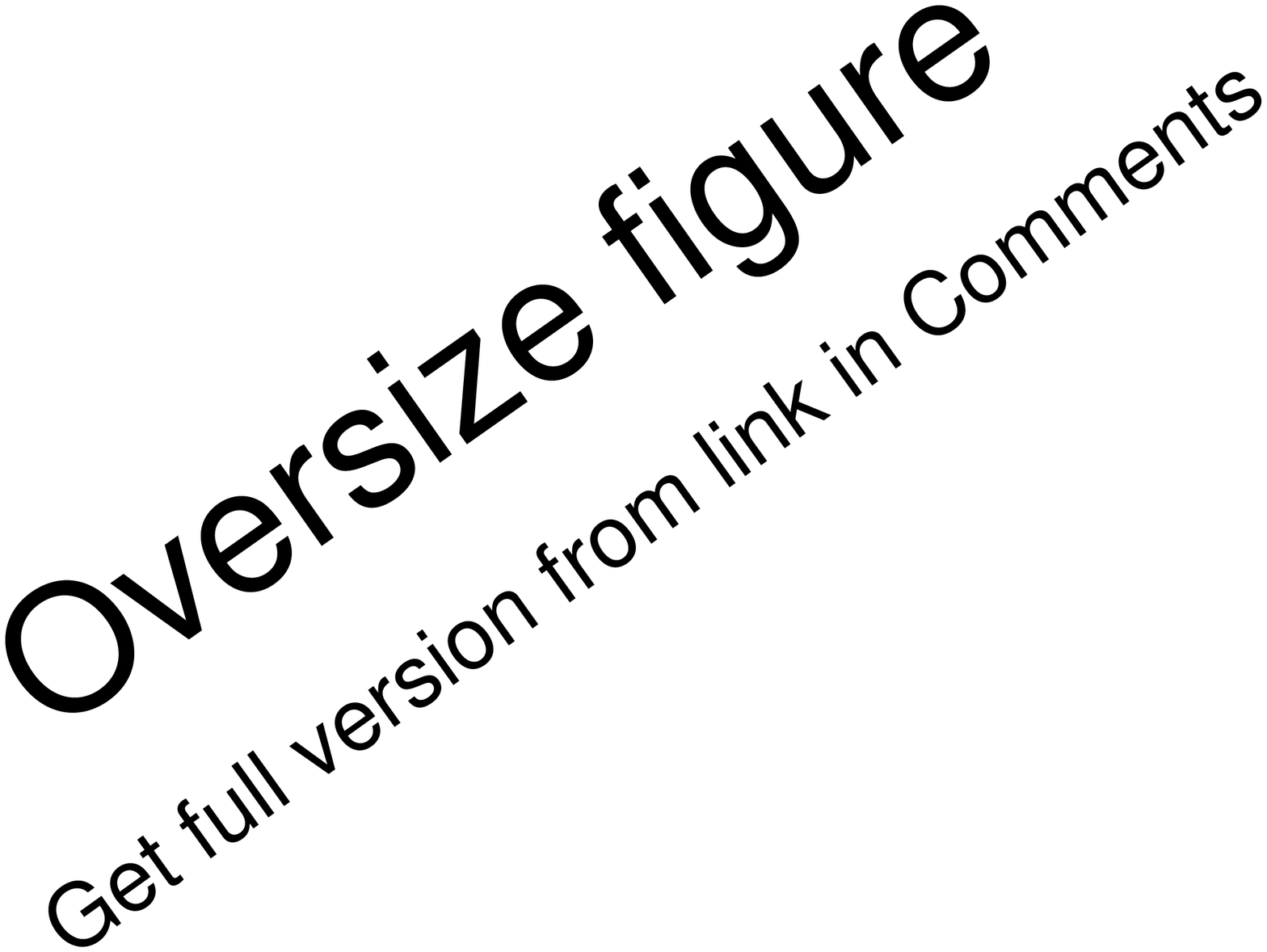}
\includegraphics[angle=-90,width=8cm]{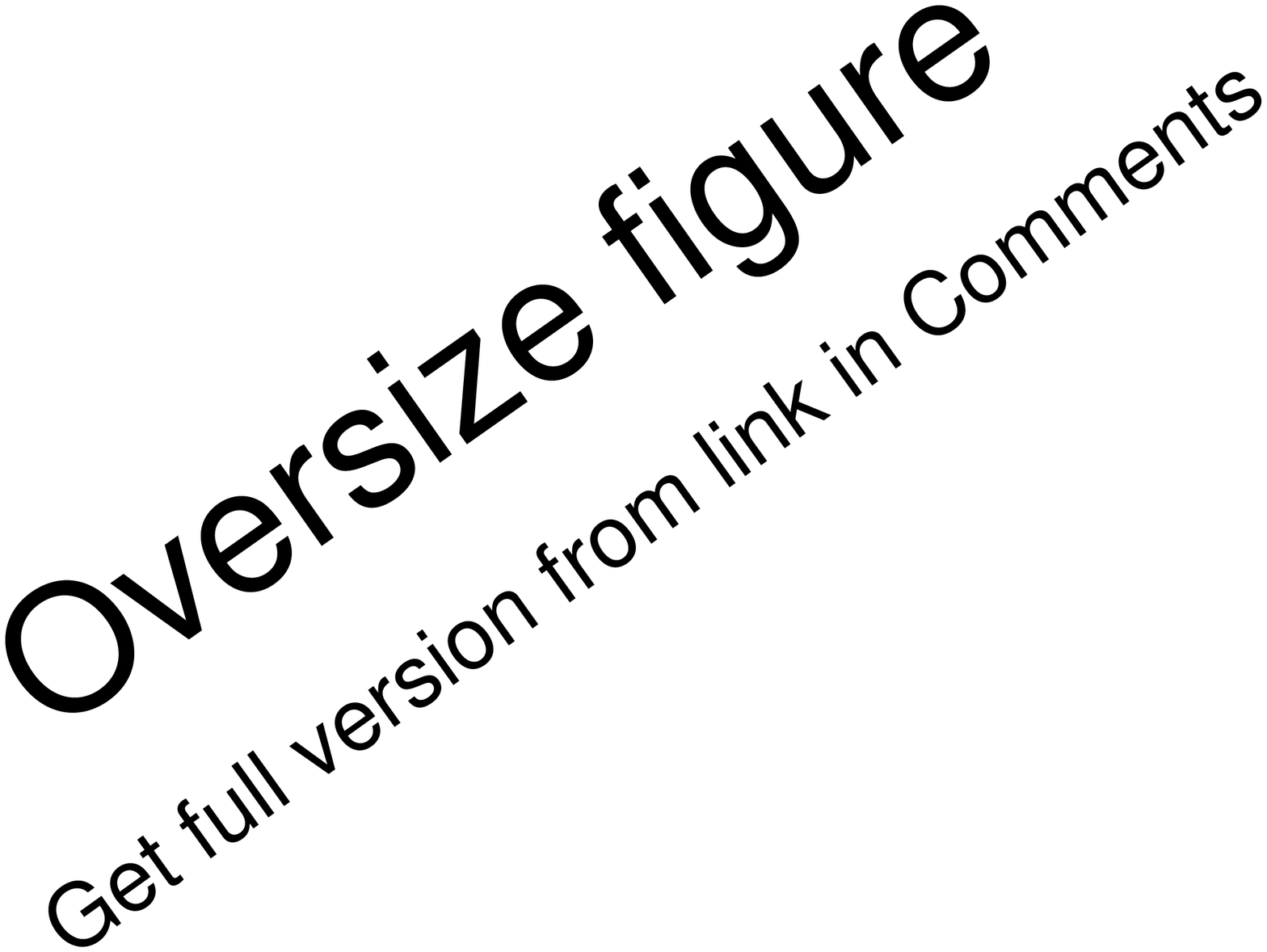}
\par\end{centering}

\caption{Magnetization switching via exponential spin-wave instability in a
box-shaped particle with uniaxial anisotropy and transverse field.
a) The particle is prepared with all spins opposite to the magnetic
field (the maximal-energy state). Again the longitudinal relaxation
is the main mechanism of switching. b) The particle is prepared with
all spins perpendicular to the magnetic field.}

\label{Fig-Uniaxial-box}
\end{figure}

\begin{figure}[th]
\begin{centering}
\includegraphics[angle=-90,width=8.5cm]{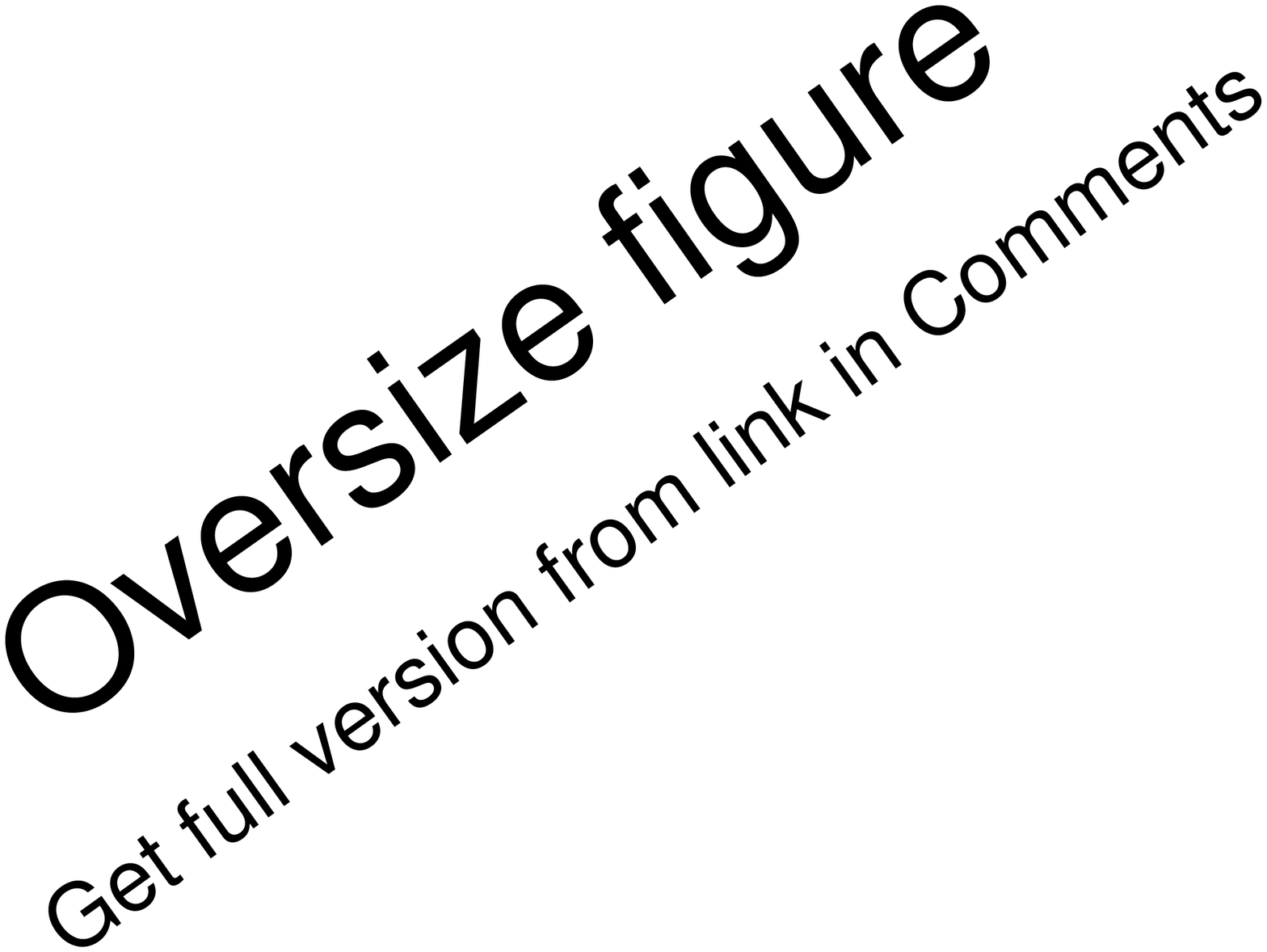}
\includegraphics[angle=-90,width=8.5cm]{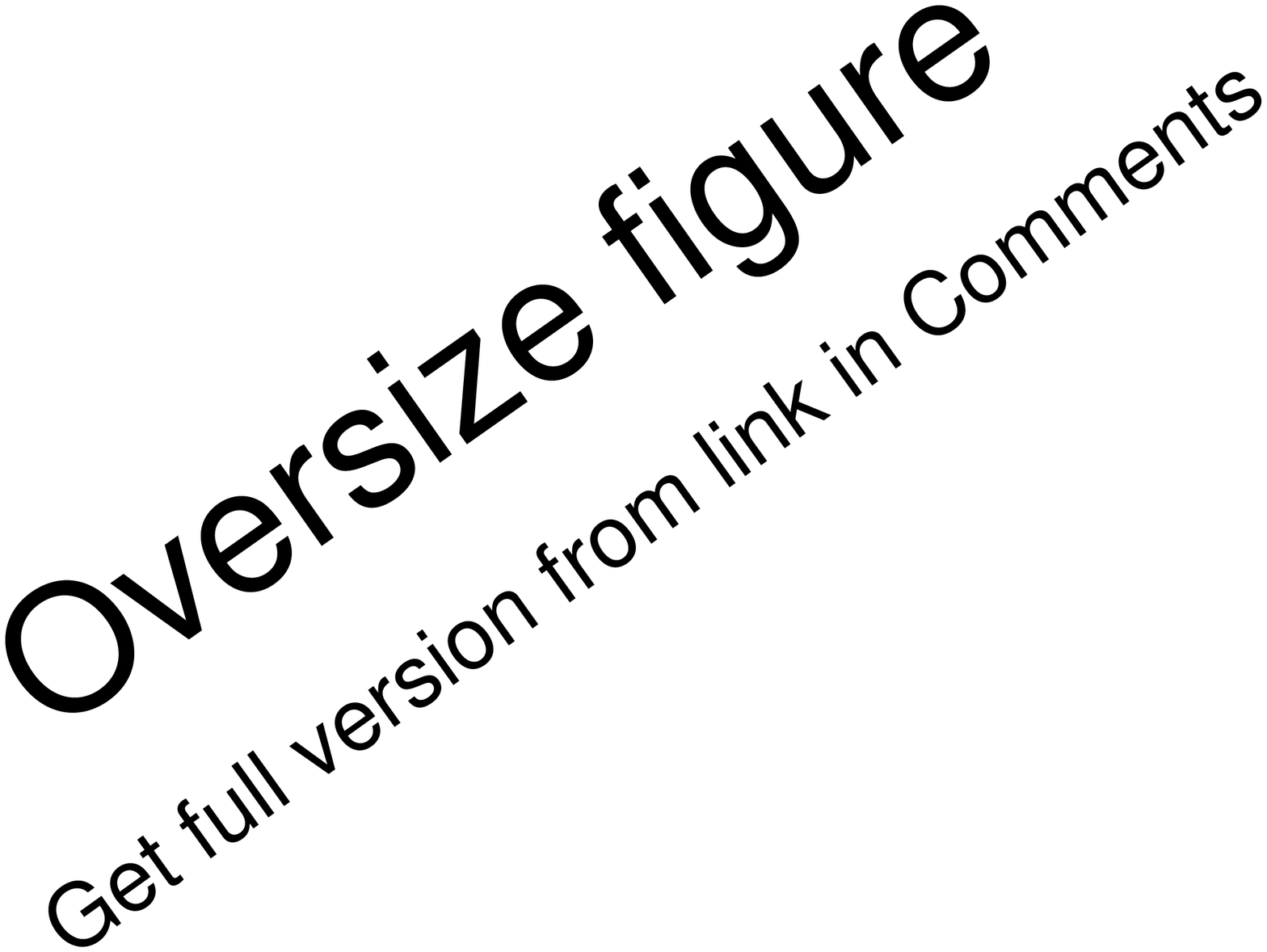}
\par\end{centering}

\caption{Magnetization switching via linear spin-wave instability in a particle
with random anisotropy. a) Particle prepared with all spins antiparallel
to the magnetic field. The small-$t$ asymptote of Eq.\ (\ref{mztinitial})
is shown by the dashed line. For this initial condition, switching
occurs predominantly via changing the magnetization length $m.$ b)
Particle prepared with all spins perpendicular to the magnetic field.}

\label{fig:Random-box}
\end{figure}

The results shown in Fig.\ \ref{fig:Uniaxial-sphere}a for the spherical
particle and in Fig. \ref{fig:Random-box}a for the box-shape particle
are similar. One can see a nearly full reversal via the \emph{longitudinal}
relaxation, since the transverse magnetization component $m_{\bot}$
remains small at all times. In the middle of the switching process,
excited spin waves almost completely destroy the magnetization $m$.
Then the first-generation long-wavelength SWs of high amplitude convert
via nonlinear processes into all possible spin wave modes and the
system thermalizes. Since the thermodynamics of classical spins is
mainly determined by short-wavelength modes of high energy, the energy
conservation requires that the amplitudes of these SWs be small. This
explains the almost full recovery of $m$ after reversal. The asymptotic
disordering $1-m>0$ corresponds to the final temperature $T$ following
from the energy balance.

The short-time magnification in Fig.\ \ref{fig:Uniaxial-sphere}b
shows an exponential increase of $1-m$ due to the SW instability
until the first dip, followed by an incomplete recovery of $m$ and
its further decrease. The same behavior was observed for other particle's
shapes.

If the initial particle's magnetization makes an angle with the highest-energy
direction, the particle's dynamics is a combination of a non-circular
precession of $\mathbf{m}$ and relaxation via SW processes, so that
our numerical results show a more complicated behavior. The closer
is the initial orientation of $\mathbf{m}$ to the energy minimum,
the slower is the relaxation towards it and the smaller is the decrease
of $m$ in the course of relaxation. The results for the box-shaped
particle with its initial magnetization vector perpendicular to the
highest-energy direction are shown in Fig. \ref{Fig-Uniaxial-box}b.
One can see that the transverse relaxation is extremely slow in this
case, so that a prohibitively long computer time is needed to follow
the decay $\left|m_{\bot}\right|\rightarrow0.$

\begin{figure}[t]
\begin{centering}
\includegraphics[angle=-90,width=10cm]{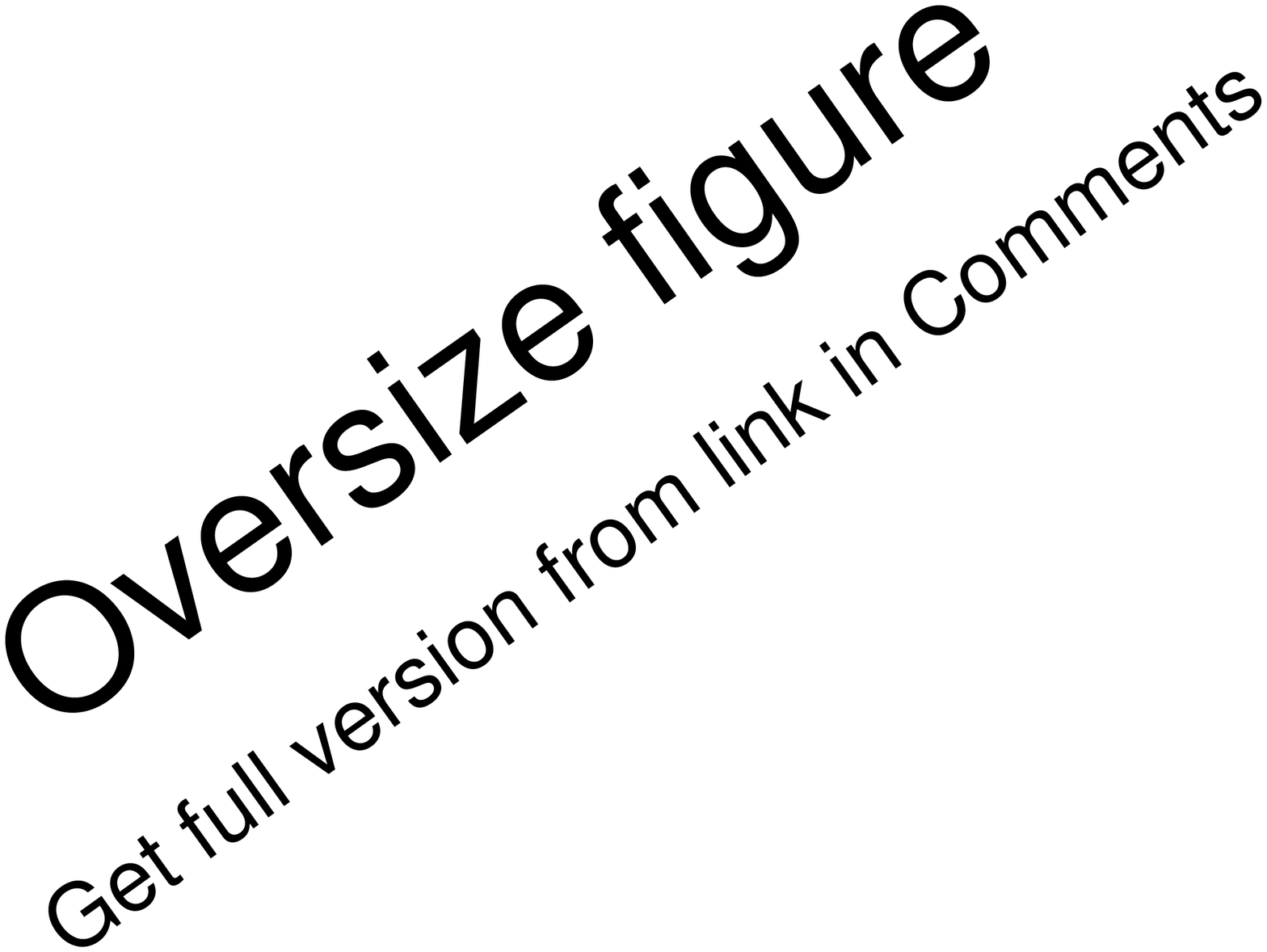}
\par\end{centering}

\caption{Magnetization switching out of the disappearing metastable state via
exponential spin-wave instability in a particle of a box shape with
uniaxial anisotropy and oblique magnetic field.}

\label{fig:UniaxialVA-oblique_field}
\end{figure}

Fig. \ref{fig:UniaxialVA-oblique_field} shows the magnetization reversal
out of a disappearing metastable state for the model with uniaxial
$z$ anisotropy and oblique magnetic field applied at the angle $\psi=\pi/4$
to the $z$ axis. For $h=D$ there is a disappearing metastable minimum
for the orientation of the spins at $\theta=3\pi/4$ to the $z$ axis.
Indeed, the energy $E(\theta)=-D\cos^{2}\theta-\left(h/\sqrt{2}\right)\cos\theta-\left(h/\sqrt{2}\right)\sin\theta,$
following from Eqs. (\ref{Ham}) and (\ref{HamAUniax}), satisfies
$E^{\prime}(\theta)=E^{\prime\prime}(\theta)=0$ for $h=D$ and $\theta=3\pi/4.$
Also one can check that for $h=D$ one has $E^{\prime}(\theta)=0$
at $\theta=\pi/12$ that is the energy minimum. Notice that for simulations,
a slightly higher value of $h$ has been chosen to let $\mathbf{m}$
precess away from the initial state with $\theta=3\pi/4.$ One can
see that this precession is strongly non-circular and the particle's
global magnetization $\mathbf{m}$ tends to return into the initial
state that would comply with the energy conservation law for a single
spin. However, due to the excitation of internal spin waves in the
particle the energy associated with $\mathbf{m}$ decreases, so that
the returning to the initial orientation is incomplete. After a couple
of cycles a lot of exponentially unstable spin waves get excited and
a faster relaxation of $\mathbf{m}$ begins. Asymptotically $\mathbf{m}$
approaches the orientation close to the energy minimum at $\theta=\pi/12,$
up to thermal disordering. The reduction of $m$ increases exponentially
at small times but does not exceed 40\% in the middle of relaxation
in this case.

Fig.\ \ref{fig:Random-box} shows the results of our numerical simulations
for the magnetization switching in the random-anisotropy model described
by Eq. (\ref{HamARand}). The particle is again of a box shape with
fbc, consisting of $28\times31\times34=29512$ atomic spins. One can
see again a nearly full reversal which is, in contrast to Fig.\ \ref{fig:Uniaxial-sphere}b,
linear at short times. The simulation results at short times are in
a reasonable accord with the analytical result in Eq. (\ref{mztinitial})
that is shown by the dashed line. Also shown in Fig.\ \ref{fig:Random-box}a
is the asymptotic approach of the thermal state with temperature $T$
corresponding to the released Zeeman energy. Note that the reduction
of $m$ in the middle of reversal is much stronger for the anticollinear
initial state (Fig.\ \ref{fig:Random-box}a) than for the perpendicular
initial state (Fig.\ \ref{fig:Random-box}b), similarly to the case
of the exponential instability.

\begin{figure}[t]
\begin{centering}
\includegraphics[angle=-90,width=10cm]{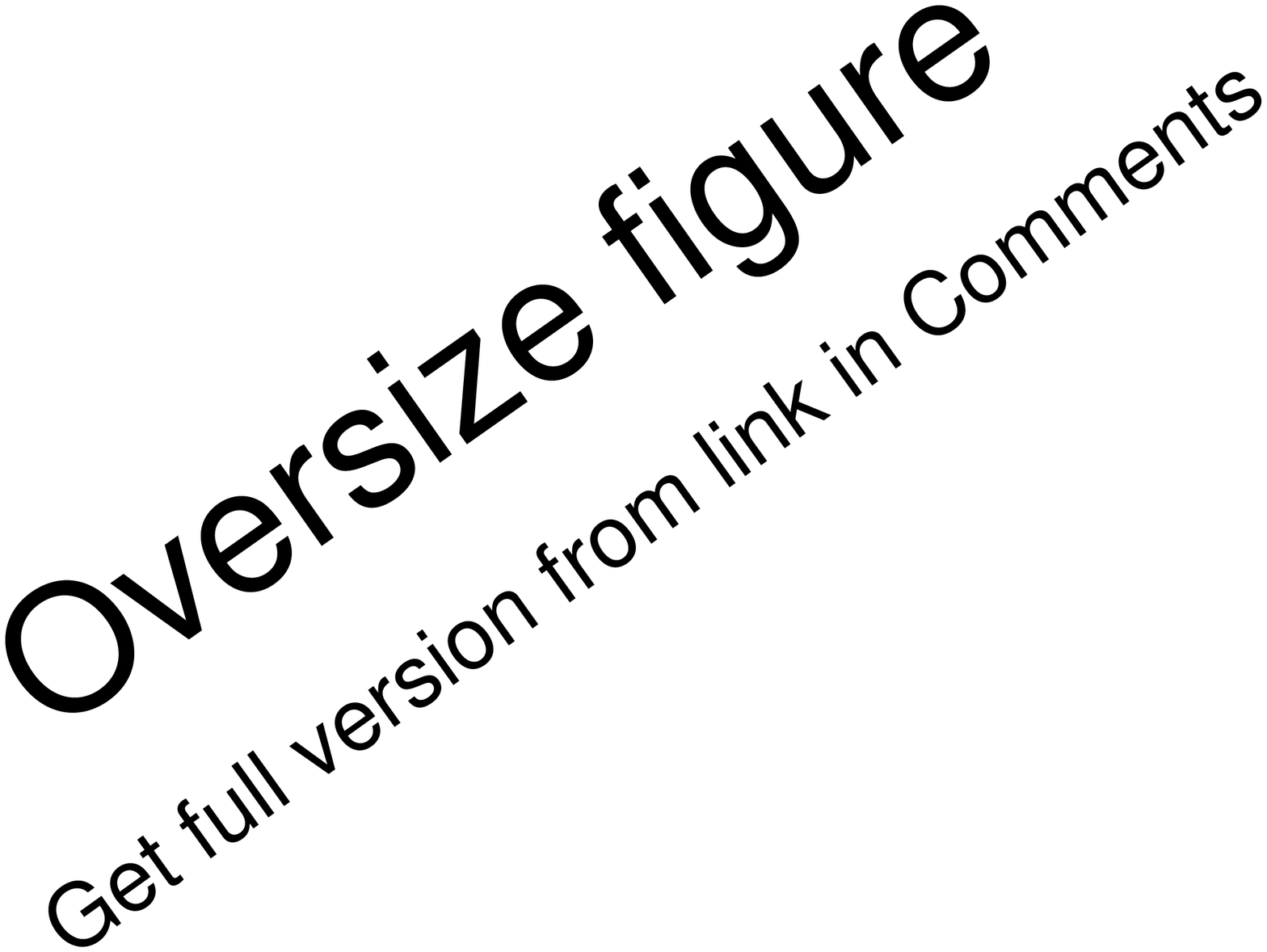}
\par\end{centering}

\caption{Time dependence of the magnetization magnitude $m$ for the particle
of a cubic shape consisting of $5\times5\times5=125$ spins with random
anisotropy. For the resonant value of the field $h$ spin waves are
generated out of the false vacuum (magnetization opposite to the field)
but they cannot be converted into other modes so that the process
is bottlenecked. For the non-resonant value of $h$ spin waves are
practically not generated.}

\label{fig:Random-box-125}
\end{figure}

As discussed in Sec. \ref{Sec-Ham}, in smaller particles SW modes
are essentially discrete and SW conversion processes are blocked by
the impossibility to satisfy the energy conservation law. Even if
one SW mode is generated as a result of exponential or linear instability,
it usually cannot be converted into other modes. This leads to the
spin-wave bottleneck in the relaxation process via internal spin waves.
As an example we consider the sc-lattice cubic particle with $N_{x}=N_{y}=N_{z}=5$
($\mathcal{N}=125$) the three degenerate modes $(1,0,0),$ $(0,1,0)$
and $(0,0,1)$ are excited via the linear instability process in the
model with random anisotropy for the magnetic fields in the vicinity
of $h/J=0.38197,$ as follows from Eqs. (\ref{ResCondSW}), (\ref{Jksc}),
and (\ref{kfbc}). In Fig. \ref{fig:Random-box-125} obtained with
$D_{R}/J=0.01$ one can see that at $h/J=0.38197$ there is an essential
reduction of the initially saturated magnetization, although the relaxation
is bottlenecked because second-generation spin waves cannot be created.
The evolution of $m$ is not sinusoidal because three degenerate SW
modes are excited with different efficiencies, depending on the realization
of the random anisotropy, so that each mode has its own conversion
frequency $\Omega,$ defined below Eq. (\ref{LinearStabCrit}). To
the contrast, for the off-resonance value $h/J=0.37$ there is no
SW mode at resonance with the false vacuum and deviations $1-m$ remain
very small.

For larger values of the random anisotropy the particle's SW spectrum
becomes effectively continuous, in accordance with the arguments at
the end of Sec. \ref{Sec-Ham}. In particular, for $D_{R}/J=0.1$
there is no principal difference between the results for $h/J=0.38197$
and $h/J=0.37$ and the magnetization relaxes towards the ground state.
However, this relaxation is much slower than in really large particles
considered above and there is a significant quasi-random dependence
of the relaxation rate on model parameters.

\section{Discussion}

\label{Sec-Disc}

We have shown that internal spin-wave processes in large enough magnetic
particles can lead to complete magnetization reversal and thermalization.
The role of thermal bath in these processes is played by the magnetic
particle itself. The energy release in the course of the relaxation
of the particle's global magnetization $\mathbf{m}$ towards the ground
state is absorbed by internal spin waves, so that the total energy
is conserved, as long as there is no coupling to the environment.

The two main scenarios of magnetization reversal are through exponential
and linear SW instabilities. While the former is the Suhl instability,
described here within a more general formalism that allows for large
magnetization motions, the linear instability is difficult to pinpoint
in the existing literature. Anyway, theoretical description of the
linear instability requires redefinition of the spin-wave vacuum by
using the frame related to the particle's global magnetization $\mathbf{m}$
that in general depends on time.

In both cases, the relaxation is fast at the beginning but then slows
down. These results are relevant to the study of the dynamics of magneto-electronic
elements and, more generally, in the physics of \emph{unstable} macroscopic
states. In the model with random anisotropy that can be solved analytically
relaxation at asymptotically large times is power law rather than
exponential. The reason is that the relaxation rate becomes small
as the direction of the particle's magnetization approaches the ground-state
direction. For the random-anisotropy model both longitudinal and transverse
relaxation rates are maximal for $\mathbf{m}$ pointing toward the
energy maximum. It is clear that the majority of the magnetic nanoparticles
studied nowadays are shown to exhibit effective uniaxial anisotropy.
However, as their size is further reduced, the on-site crystalline
anisotropy may adopt a more disordered distribution upon which the
model of random anisotropy may become of some relevance.

Estimations of relaxation rates via internal spin waves in metallic
Co made in Ref. \onlinecite{garkacrey08epl} yield $\Gamma\sim$10$^{5}-10^{6}$
s$^{-1}.$ Comparison with the spin-phonon and other external rates,
obtained microscopically, is difficult since the latter are still
unreliable and should be modified by collective processes such as
phonon bottleneck\cite{garanin07prb,garanin08prb} and phonon superradiance.\cite{chugar04prl}
In most cases internal processes should dominate. One should not forget,
however, that the latter die out for nanoparticles that cannot accommodate
spin waves. The phenomenological Landau-Lifshitz relaxation rate $\Gamma_{LL}=\alpha h/\hbar$
with the damping constant $\alpha$ ranging between $10^{-1}$ and
$10^{-3}$ is typically much larger than microscopic rates that do
not take into account collective processes. One has to be careful,
however, since the $\alpha$ that is extracted from experiments may
contain a contribution from the SW processes.

Numerical simulations in this work have been done for a classical-spin
model without coupling to the environment. The dynamical equations
are Larmor equations for each atomic spin precessing in the effective
field created by other spins plus the magnetic and anisotropy field.
Including the effect of the environment by a double vector product
relaxation term introduced by Landau and Lifshitz\cite{lanlif35}
is straightforward. Also one can include the finite-temperature effect
by the Langevin fields that will slow down the program, however. It
should be stressed that such a method of including the environment
is not reliable because it misses collective effects in the spin-lattice
relaxation.

\section{Acknowledgements}

D.G. thanks E. M. Chudnovsky for useful discussions. This work has
been supported by the Cottrell College Science Award of the
Research Corporation. We are greatly indebted to L. Reynaud for
his expert support in questions of numerical efficiency.

\newpage{}

\appendix
%dummy comment inserted by tex2lyx to ensure that this paragraph is not empty

\section{Random anisotropy: Dynamics in the precessing frame}

Consider a model with $\mathcal{H}_{Ai}=0$ and small random anisotropy
$\overleftrightarrow{\mathbf{g}}_{i}$ that is the only source for
the particle's relaxation. In the zeroth approximation $\mathbf{m}$
is simply precessing around the magnetic field, \begin{equation}
\hbar\mathbf{\dot{m}}=\left[\mathbf{m\times h}\right],\label{mPrecess}\end{equation}
 and it is convenient to consider spin waves in the frame related
to $\mathbf{m}$ and defined by Eq. (\ref{RotFrameDef}). Eqs. (\ref{RotFrameDef})
can be solved for $\mathbf{h}$ resulting in \begin{equation}
\mathbf{h=}hx\mathbf{n-}h\sqrt{1-x^{2}}\mathbf{e}_{2}\label{hProjections}\end{equation}
 Time derivatives of the basis vectors defined by Eq. (\ref{RotFrameDef})
are given by (using (\ref{mPrecess})) \begin{eqnarray}
\hbar\mathbf{\dot{n}} & = & \left[\mathbf{n}\times\mathbf{h}\right]\notag\\
\hbar\mathbf{\dot{e}}_{1} & = & \frac{\left[\hbar\mathbf{\dot{n}\times h}\right]}{h\sqrt{1-x^{2}}}=\frac{\left[\left[\mathbf{n}\times\mathbf{h}\right]\times\mathbf{h}\right]}{h\sqrt{1-x^{2}}}=\frac{-\mathbf{n}h+\mathbf{h}x}{\sqrt{1-x^{2}}}\notag\\
\hbar\mathbf{\dot{e}}_{2} & = & \frac{\hbar\mathbf{\dot{n}}\left(\mathbf{n}\cdot\mathbf{h}\right)}{h\sqrt{1-x^{2}}}=\frac{\left[\mathbf{n}\times\mathbf{h}\right]x}{\sqrt{1-x^{2}}}\end{eqnarray}
 and they can be projected on $\mathbf{n},$ $\mathbf{e}_{1},$ and
$\mathbf{e}_{2}:$\begin{eqnarray}
\hbar\mathbf{\dot{n}} & = & h\sqrt{1-x^{2}}\mathbf{e}_{1}\notag\\
\hbar\mathbf{\dot{e}}_{1} & = & \left(\mathbf{n}\cdot\hbar\mathbf{\dot{e}}_{1}\right)\mathbf{n}+\left(\mathbf{e}_{2}\cdot\hbar\mathbf{\dot{e}}_{1}\right)\mathbf{e}_{2}=-h\sqrt{1-x^{2}}\mathbf{n}-hx\mathbf{e}_{2}\notag\\
\hbar\mathbf{\dot{e}}_{2} & = & hx\mathbf{e}_{1}.\label{e12Ders}\end{eqnarray}
 Now the explicit time dependence have to be obtained. To this end,
it is convenient to choose the $z$ axis along $\mathbf{h},$ i.e.,
$\mathbf{h}=h\mathbf{e}_{z},$ then from Eq. (\ref{mPrecess}) follows
\begin{eqnarray}
\mathbf{n}(t) & = & x\mathbf{e}_{z}\mathbf{-}\sqrt{1-x^{2}}\cos(\omega_{h}t)\mathbf{e}_{x}+\sqrt{1-x^{2}}\sin(\omega_{h}t)\mathbf{e}_{y}\notag\\
\mathbf{e}_{1}(t) & = & \cos(\omega_{h}t)\mathbf{e}_{y}+\sin(\omega_{h}t)\mathbf{e}_{x}\notag\\
\mathbf{e}_{2}(t) & = & -\sqrt{1-x^{2}}\mathbf{e}_{z}\mathbf{-}x\cos(\omega_{h}t)\mathbf{e}_{x}+x\sin(\omega_{h}t)\mathbf{e}_{y}.\label{ne12tDep}\end{eqnarray}
 where \begin{equation}
\omega_{h}=h/\hbar.\label{omegahDef}\end{equation}

One can project $\mathbf{\psi}_{i}$ onto the time dependent $\mathbf{n},$
$\mathbf{e}_{1}$ and $\mathbf{e}_{2}$ as \begin{equation}
\mathbf{\psi}_{i}\mathbf{=}\psi_{i0}\mathbf{n}+\psi_{i1}\mathbf{e}_{1}+\psi_{i2}\mathbf{e}_{2},\label{psiRotFrame}\end{equation}
 and thus \begin{eqnarray}
\hbar\mathbf{\dot{\psi}}_{i} & = & \hbar\dot{\psi}_{i0}\mathbf{n+}\hbar\dot{\psi}_{i1}\mathbf{e}_{1}+\hbar\dot{\psi}_{i2}\mathbf{e}_{2}\notag\\
 &  & {}+\psi_{i0}\hbar\mathbf{\dot{n}+}\psi_{i1}\hbar\mathbf{\dot{e}}_{1}+\psi_{i2}\hbar\mathbf{\dot{e}}_{2}.\label{psiDer}\end{eqnarray}
 Inserting this form into Eq. (\ref{psiiEq}) with $\mathcal{H}_{Ai}=0$
and $\mathbf{A}_{i}^{(2)}$ dropped and projecting onto the three
basis vectors one obtains the equation \begin{equation}
\hbar\dot{\psi}_{i0}-\psi_{i1}h\sqrt{1-x^{2}}=\mathbf{n}\cdot\mathbf{A}_{i}^{(1)}\label{psiioEq}\end{equation}
 as well as \begin{eqnarray}
 &  & \hbar\dot{\psi}_{i1}+\psi_{i0}h\sqrt{1-x^{2}}+hx\psi_{i2}\notag\\
 & = & \mathbf{e}_{1}\cdot\left[\mathbf{m\times}2\overleftrightarrow{\mathbf{g}}_{i}\mathbf{m}\right]+\mathbf{e}_{1}\cdot\mathbf{A}_{i}^{(1)}\label{psii1Eq}\end{eqnarray}
 together with \begin{equation}
\hbar\dot{\psi}_{i2}-hx\psi_{i1}=\mathbf{e}_{2}\cdot\left[\mathbf{m\times}2\overleftrightarrow{\mathbf{g}}_{i}\mathbf{m}\right]+\mathbf{e}_{2}\cdot\mathbf{A}_{i}^{(1)}.\label{psii2Eq}\end{equation}
 In the rhs of Eq. (\ref{psiioEq}) one has \begin{equation}
\mathbf{n}\cdot\mathbf{A}_{i}^{(1)}=\mathbf{n}\cdot\left[\mathbf{\psi}_{i}\times\mathbf{h}\right]=\mathbf{\psi}_{i}\cdot\left[\mathbf{h}\times\mathbf{n}\right].\end{equation}

Then with the help of Eq. (\ref{RotFrameDef}) one obtains \begin{equation}
\dot{\psi}_{i0}=0.\label{psii0dotEq}\end{equation}
 Expressions in the rhs of other equations can be processed as follows
\begin{eqnarray}
 &  & \mathbf{e}_{1}\cdot\left[\mathbf{m\times}\overleftrightarrow{\mathbf{g}}_{i}\mathbf{m}\right]\notag\\
 & = & m^{2}\mathbf{e}_{1}\cdot\left[\mathbf{n\times}\left\{ \mathbf{e}_{1}\left(\mathbf{e}_{1}\overleftrightarrow{\mathbf{g}}_{i}\mathbf{n}\right)+\mathbf{e}_{2}\left(\mathbf{e}_{2}\overleftrightarrow{\mathbf{g}}_{i}\mathbf{n}\right)\right\} \right]\notag\\
 & = & m^{2}\mathbf{e}_{1}\cdot\left\{ \mathbf{e}_{2}\left(\mathbf{e}_{1}\overleftrightarrow{\mathbf{g}}_{i}\mathbf{n}\right)-\mathbf{e}_{1}\left(\mathbf{e}_{2}\overleftrightarrow{\mathbf{g}}_{i}\mathbf{n}\right)\right\} \notag\\
 & = & -m^{2}\left(\mathbf{e}_{2}\overleftrightarrow{\mathbf{g}}_{i}\mathbf{n}\right)\label{transformation}\end{eqnarray}
 and, similarly, \begin{equation}
\mathbf{e}_{2}\cdot\left[\mathbf{m\times}\overleftrightarrow{\mathbf{g}}_{i}\mathbf{m}\right]=m^{2}\left(\mathbf{e}_{1}\overleftrightarrow{\mathbf{g}}_{i}\mathbf{n}\right).\end{equation}
 From Eq. (\ref{A1iDef}) with $\mathcal{H}_{Ai}=0$ one obtains \begin{eqnarray}
\mathbf{A}_{i}^{(1)} & = & \left[\mathbf{\psi}_{i}\times\left(\mathbf{h}+J_{0}\mathbf{m}\right)\right]+\left[\mathbf{m\times}\sum_{j}J_{ij}\mathbf{\psi}_{j}\right]\notag\\
 & = & \left[\left(\psi_{i0}\mathbf{n}+\psi_{i1}\mathbf{e}_{1}+\psi_{i2}\mathbf{e}_{2}\right)\times\left(\mathbf{h}+J_{0}\mathbf{m}\right)\right]\notag\\
 &  & +\left[\mathbf{m\times}\sum_{j}J_{ij}\left(\psi_{j0}\mathbf{n}+\psi_{j1}\mathbf{e}_{1}+\psi_{j2}\mathbf{e}_{2}\right)\right]\notag\\
 & = & \psi_{i0}\left[\mathbf{n}\times\mathbf{h}\right]+\psi_{i1}\left(\left[\mathbf{e}_{1}\times\mathbf{h}\right]+J_{0}m\left[\mathbf{e}_{1}\times\mathbf{n}\right]\right)\notag\\
 &  & \qquad+\psi_{i2}\left(\left[\mathbf{e}_{2}\times\mathbf{h}\right]+J_{0}m\left[\mathbf{e}_{2}\times\mathbf{n}\right]\right)\notag\\
 &  & +m\sum_{j}J_{ij}\left(\psi_{j1}\left[\mathbf{n\times e}_{1}\right]+\psi_{j2}\left[\mathbf{n\times e}_{2}\right]\right).\end{eqnarray}
 With the help of Eqs. (\ref{RotFrameDef}) and (\ref{hProjections})
this becomes \begin{eqnarray}
\mathbf{A}_{i}^{(1)} & = & \psi_{i0}h\sqrt{1-x^{2}}\mathbf{e}_{1}\notag\\
 &  & +\psi_{i1}\left(-hx\mathbf{e}_{2}\mathbf{-}h\sqrt{1-x^{2}}\mathbf{n}-J_{0}m\mathbf{e}_{2}\right)\notag\\
 &  & +\psi_{i2}\left(hx\mathbf{e}_{1}+J_{0}m\mathbf{e}_{1}\right)\notag\\
 &  & +m\sum_{j}J_{ij}\left(\psi_{j1}\mathbf{e}_{2}-\psi_{j2}\mathbf{e}_{1}\right)\end{eqnarray}
 whose components of this vector are \begin{eqnarray}
A_{i1}^{(1)} & = & hx\psi_{i2}+m\sum_{j}\left(J_{0}\delta_{ij}-J_{ij}\right)\psi_{j2}\notag\\
 &  & {}\qquad+h\sqrt{1-x^{2}}\psi_{i0}\notag\\
A_{i2}^{(1)} & = & -hx\psi_{i1}-m\sum_{j}\left(J_{0}\delta_{ij}-J_{ij}\right)\psi_{j1.}\end{eqnarray}
 Now Eqs. (\ref{psii1Eq}) and (\ref{psii2Eq}) after simplification
become \begin{eqnarray}
\hbar\dot{\psi}_{i1} & = & m\sum_{j}\left(J_{0}\delta_{ij}-J_{ij}\right)\psi_{j2}-2m^{2}\left(\mathbf{e}_{2}\overleftrightarrow{\mathbf{g}}_{i}\mathbf{n}\right)\notag\\
\hbar\dot{\psi}_{i2} & = & -m\sum_{j}\left(J_{0}\delta_{ij}-J_{ij}\right)\psi_{j1}+2m^{2}\left(\mathbf{e}_{1}\overleftrightarrow{\mathbf{g}}_{i}\mathbf{n}\right).\notag\end{eqnarray}
 This is the system of equations describing the generation of internal
spin waves in a magnetic particle.

It is convenient to introduce the variables \begin{equation}
\psi_{i,\pm}=\psi_{i1}\pm i\psi_{i2},\qquad\mathbf{e}_{\pm}\equiv\mathbf{e}_{1}\pm i\mathbf{e}_{2}\label{zetakDef}\end{equation}
 so that \begin{eqnarray}
\mathbf{e}_{1} & = & \frac{1}{2}\left(\mathbf{e}_{-}+\mathbf{e}_{+}\right),\qquad\mathbf{e}_{2}=\frac{i}{2}\left(\mathbf{e}_{-}-\mathbf{e}_{+}\right)\notag\\
\psi_{i1} & = & \frac{1}{2}\left(\psi_{i,-}+\psi_{i,+}\right),\qquad\psi_{i2}=\frac{i}{2}\left(\psi_{i,-}-\psi_{i,+}\right)\notag\end{eqnarray}
 and \begin{eqnarray}
\mathbf{e}_{1}\psi_{i1}+\mathbf{e}_{2}\psi_{i2} & = & Re\left(\mathbf{e}_{-}\psi_{i,+}\right)\notag\\
\mathbf{e}_{1}\psi_{i2}-\mathbf{e}_{2}\psi_{i1} & = & Im\left(\mathbf{e}_{-}\psi_{i,+}\right)\notag\\
\mathbf{e}_{1}\psi_{i2}+\mathbf{e}_{2}\psi_{i1} & = & Im\left(\mathbf{e}_{+}\psi_{i,+}\right).\label{psieCombinations}\end{eqnarray}
 Then one obtains the equation \begin{equation}
\hbar\dot{\psi}_{i,+}=-im\sum_{j}\left(J_{0}\delta_{ij}-J_{ij}\right)\psi_{j,+}+2im^{2}\left(\mathbf{e}_{+}\overleftrightarrow{\mathbf{g}}_{i}\mathbf{n}\right).\label{zetaiEq}\end{equation}
\textbf{what about $\psi_{i,-}$?}

For a box-shaped particle one can rewrite this equation in terms of
the discrete Fourier components \begin{equation}
\mathbf{\psi}_{\mathbf{k}}=\sum_{i}e^{i\mathbf{k\cdot r}_{i}}\mathbf{\psi}_{i},\qquad\mathbf{\psi}_{i}=\frac{1}{\mathcal{N}}\sum_{\mathbf{k}}e^{-i\mathbf{k\cdot r}_{i}}\mathbf{\psi}_{\mathbf{k}}\end{equation}
 etc., as \begin{equation}
\hbar\dot{\psi}_{\mathbf{k,+}}=-i\varepsilon_{\mathrm{ex},\mathbf{k}}\psi_{\mathbf{k,+}}+2im^{2}\left(\mathbf{e}_{+}\overleftrightarrow{\mathbf{g}}_{\mathbf{k}}\mathbf{n}\right),\label{zetakEq}\end{equation}
 where \begin{equation}
\varepsilon_{\mathrm{ex},\mathbf{k}}=m\left(J_{0}-J_{\mathbf{k}}\right).\label{epskDef}\end{equation}
 The density of pure-exchange spin-wave states is given by \begin{equation}
\rho_{\mathrm{ex}}(\omega)=\frac{1}{\mathcal{N}}\sum_{\mathbf{k}}\delta\left(\omega_{\mathrm{ex},\mathbf{k}}-\omega\right),\label{rhoDef}\end{equation}
 where $\hbar\omega_{\mathrm{ex},\mathbf{k}}=\varepsilon_{\mathrm{ex},\mathbf{k}}$
and $\mathcal{N}=N_{x}N_{y}N_{z}$. It satisfies $\int d\omega\rho(\omega)=1.$
In the continuous approximation for small wave vectors for the sc
lattice one has \begin{equation}
\varepsilon_{\mathrm{ex},\mathbf{k}}\cong mJ\left(ak\right)^{2}\label{epsExksmall}\end{equation}
 for a particle of a box shape with fbc for $\hbar\omega\ll J_{0}$
one has \begin{eqnarray}
\rho_{\mathrm{ex}}(\omega) & \cong & a^{3}\int\int\int_{0}^{\infty}\frac{dk_{x}dk_{y}dk_{z}}{\pi^{3}}\delta\left(\omega_{\mathrm{ex},\mathbf{k}}-\omega\right)\notag\\
 & = & \hbar\frac{4\pi a^{3}}{8\pi^{3}}\int_{0}^{\infty}k^{2}dk\,\delta\left(\varepsilon_{\mathrm{ex},\mathbf{k}}-\varepsilon\right)\notag\\
 & = & \frac{\hbar}{\left(2\pi\right)^{2}}\sqrt{\frac{\varepsilon}{m^{3}J^{3}}}.\label{rhoeps}\end{eqnarray}
 Eq. (\ref{zetakEq}) has the solution \begin{eqnarray}
 &  & \psi_{\mathbf{k,+}}(t)=\psi_{\mathbf{k,+}}(0)e^{-i\omega_{\mathrm{ex},\mathbf{k}}t}+2m^{2}\frac{i}{\hbar}\int_{0}^{t}dt^{\prime}\notag\\
 &  & \times e^{-i\omega_{\mathrm{ex},\mathbf{k}}(t-t^{\prime})}\left(\mathbf{e}_{+}(t^{\prime})\overleftrightarrow{\mathbf{g}}_{\mathbf{k}}\mathbf{n}(t^{\prime})\right),\label{zetakSol}\end{eqnarray}
where $\hbar\omega_{\mathrm{ex},\mathbf{k}}=\varepsilon_{\mathrm{ex},\mathbf{k}}.$
The first term of the solution takes into account spin waves already
available in the particle, such as thermal spin waves. The second
term describes spin waves generated by the particle's precession via
the random anisotropy. Note that, according to Eq. (\ref{psii0dotEq}),
longitudinal fluctuations $\psi_{i0}$ do not have any dynamics in
the linear approximation.

The next step is to substitute the solution for $\psi_{\mathbf{k,+}}(t)$
into the relaxation term $\mathbf{R}$ in the equation for $\mathbf{m},$
Eq. (\ref{mEq0}). Keeping terms of order $\mathbf{\psi g}$ in Eq.
(\ref{RDef}) with $\mathcal{H}_{Ai}=0$ one obtains \begin{equation}
\mathbf{R}=2m\frac{1}{\mathcal{N}}\sum_{i}\left\{ \left[\mathbf{n\times}\overleftrightarrow{\mathbf{g}}_{i}\mathbf{\psi}_{i}\right]+\left[\mathbf{\psi}_{i}\times\overleftrightarrow{\mathbf{g}}_{i}\mathbf{n}\right]\right\} .\end{equation}
 For $R_{\Vert}$ of Eq. (\ref{RparperpDef}), using transformations
in Eq. (\ref{transformation}) one obtains \begin{eqnarray}
R_{\Vert} & = & 2m\frac{1}{\mathcal{N}}\sum_{i}\mathbf{n\cdot}\left[\mathbf{\psi}_{i}\times\overleftrightarrow{\mathbf{g}}_{i}\mathbf{n}\right]\notag\\
 & = & -2m\frac{1}{\mathcal{N}}\sum_{i}\mathbf{\psi}_{i}\mathbf{\cdot}\left[\mathbf{n}\times\overleftrightarrow{\mathbf{g}}_{i}\mathbf{n}\right]\notag\\
 & = & -2m\frac{1}{\mathcal{N}}\sum_{i}\mathbf{\psi}_{i}\mathbf{\cdot}\left[\mathbf{n}\times\left(\mathbf{e}_{1}\left(\mathbf{e}_{1}\overleftrightarrow{\mathbf{g}}_{i}\mathbf{n}\right)+\mathbf{e}_{2}\left(\mathbf{e}_{2}\overleftrightarrow{\mathbf{g}}_{i}\mathbf{n}\right)\right)\right]\notag\\
 & = & -2m\frac{1}{\mathcal{N}}\sum_{i}\mathbf{\psi}_{i}\mathbf{\cdot}\left\{ \mathbf{e}_{2}\left(\mathbf{e}_{1}\overleftrightarrow{\mathbf{g}}_{i}\mathbf{n}\right)-\mathbf{e}_{1}\left(\mathbf{e}_{2}\overleftrightarrow{\mathbf{g}}_{i}\mathbf{n}\right)\right\} \notag\\
 & = & 2m\frac{1}{\mathcal{N}}\sum_{i}\left\{ \left(\mathbf{e}_{2}\overleftrightarrow{\mathbf{g}}_{i}\mathbf{n}\right)\psi_{i1}-\left(\mathbf{e}_{1}\overleftrightarrow{\mathbf{g}}_{i}\mathbf{n}\right)\psi_{i2}\right\} .\end{eqnarray}
 Finally, with the help of Eq. (\ref{psieCombinations}) \begin{equation}
R_{\Vert}=-2m\, Im\left[\frac{1}{\mathcal{N}}\sum_{i}\left(\mathbf{e}_{-}\overleftrightarrow{\mathbf{g}}_{i}\mathbf{n}\right)\psi_{i,+}\right].\label{RparFin}\end{equation}
 Further one obtains \begin{eqnarray}
R_{2} & = & \mathbf{e}_{2}\mathbf{\cdot R=}2m\frac{1}{\mathcal{N}}\sum_{i}\mathbf{e}_{2}\mathbf{\cdot}\left\{ \left[\mathbf{n\times}\overleftrightarrow{\mathbf{g}}_{i}\mathbf{\psi}_{i}\right]+\left[\mathbf{\psi}_{i}\times\overleftrightarrow{\mathbf{g}}_{i}\mathbf{n}\right]\right\} \notag\\
 & = & 2m\frac{1}{\mathcal{N}}\sum_{i}\left\{ \left(\mathbf{e}_{1}\overleftrightarrow{\mathbf{g}}_{i}\mathbf{\psi}_{i}\right)-\mathbf{\psi}_{i}\cdot\left[\mathbf{e}_{2}\times\overleftrightarrow{\mathbf{g}}_{i}\mathbf{n}\right]\right\} \notag\\
 & = & 2m\frac{1}{\mathcal{N}}\sum_{i}\left\{ \left(\mathbf{e}_{1}\overleftrightarrow{\mathbf{g}}_{i}\mathbf{\psi}_{i}\right)\right.\notag\\
 &  & \qquad-\left.\mathbf{\psi}_{i}\cdot\left(\mathbf{e}_{1}\left(\mathbf{n}\overleftrightarrow{\mathbf{g}}_{i}\mathbf{n}\right)-\mathbf{n}\left(\mathbf{e}_{1}\overleftrightarrow{\mathbf{g}}_{i}\mathbf{n}\right)\right)\right\} \notag\\
 & = & 2m\frac{1}{\mathcal{N}}\sum_{i}\left\{ \left(\mathbf{e}_{2}\overleftrightarrow{\mathbf{g}}_{i}\mathbf{e}_{1}\right)\psi_{i1}+\left(\mathbf{e}_{2}\overleftrightarrow{\mathbf{g}}_{i}\mathbf{e}_{2}\right)\psi_{i2}\right.\notag\\
 &  & \qquad-\left.\psi_{i1}\left(\mathbf{n}\overleftrightarrow{\mathbf{g}}_{i}\mathbf{n}\right)\right\} ,\end{eqnarray}
 where we have dropped all the terms with $\psi_{i0}$ since they
are frozen in and disappear after the averaging over the random anisotropy.
This can be rewritten as \begin{equation}
R_{2}=2m\, Re\left[\frac{1}{\mathcal{N}}\sum_{i}\left\{ \left(\mathbf{e}_{1}\overleftrightarrow{\mathbf{g}}_{i}\mathbf{e}_{-}\right)-\left(\mathbf{n}\overleftrightarrow{\mathbf{g}}_{i}\mathbf{n}\right)\right\} \mathbf{\psi}_{i,+}\right].\label{R2Final}\end{equation}

In the Fourier representation $R_{\Vert}$ and $R_{2}\equiv R_{\bot}$
are given by \begin{eqnarray}
R_{\Vert} & = & -2m\, Im\left[\frac{1}{\mathcal{N}^{2}}\sum_{\mathbf{k}}\left(\mathbf{e}_{-}\overleftrightarrow{\mathbf{g}}_{-\mathbf{k}}\mathbf{n}\right)\psi_{\mathbf{k},+}\right]\notag\\
R_{\bot} & = & 2m\, Re\left[\frac{1}{\mathcal{N}^{2}}\sum_{\mathbf{k}}\right.\notag\\
 &  & \times\left.\left\{ \left(\mathbf{e}_{1}\overleftrightarrow{\mathbf{g}}_{-\mathbf{k}}\mathbf{e}_{-}\right)-\left(\mathbf{n}\overleftrightarrow{\mathbf{g}}_{-\mathbf{k}}\mathbf{n}\right)\right\} \psi_{\mathbf{k},+}\right].\label{RCompsFourier}\end{eqnarray}
 Note the relations \begin{equation}
\overleftrightarrow{\mathbf{g}}_{\mathbf{k}}^{\ast}=\overleftrightarrow{\mathbf{g}}_{-\mathbf{k}},\qquad\psi_{\mathbf{k},\pm}^{\ast}=\psi_{-\mathbf{k},\mp}.\end{equation}
 Using Eq. (\ref{zetakSol}) yields \begin{eqnarray}
R_{\Vert} & = & -2m^{3}\, Re\left[\frac{1}{\mathcal{N}^{2}}\sum_{\mathbf{k}}\frac{1}{\hbar}\int_{0}^{t}dt^{\prime}e^{-i\omega_{\mathrm{ex},\mathbf{k}}(t-t^{\prime})}\right.\notag\\
 &  & \times\left.\left(\mathbf{e}_{-}(t)\overleftrightarrow{\mathbf{g}}_{-\mathbf{k}}\mathbf{n}(t)\right)\left(\mathbf{e}_{+}(t^{\prime})\overleftrightarrow{\mathbf{g}}_{\mathbf{k}}\mathbf{n}(t^{\prime})\right)\right].\end{eqnarray}
 Random anisotropy has the form of Eq. (\ref{gRADef}), so that \begin{equation}
g_{\mathbf{k},\alpha\beta}=D_{R}\sum_{i}e^{i\mathbf{k\cdot r}_{i}}u_{i\alpha}u_{i\beta}.\label{Falbet}\end{equation}
 Using the formula for the ensemble averaging on a lattice site $i$\begin{eqnarray}
 &  & \left\langle \left(u_{\alpha}u_{\beta}-\frac{1}{3}\delta_{\alpha\beta}\right)\left(u_{\gamma}u_{\delta}-\frac{1}{3}\delta_{\gamma\delta}\right)\right\rangle \notag\\
 & = & \left\langle u_{\alpha}u_{\beta}u_{\gamma}u_{\delta}\right\rangle +\frac{1}{9}\delta_{\alpha\beta}\delta_{\gamma\delta}\notag\\
 &  & {}+\left(\delta_{\alpha\beta}\delta_{\gamma\delta}+\delta_{\alpha\gamma}\delta_{\beta\delta}+\delta_{\alpha\delta}\delta_{\beta\gamma}\right)\left(1-\delta_{\alpha\beta\gamma\delta}\right)\frac{1}{15}\notag\\
 &  & {}+\delta_{\alpha\beta\gamma\delta}\frac{1}{5}+\frac{1}{9}\delta_{\alpha\beta}\delta_{\gamma\delta}\notag\\
 & = & \frac{1}{15}\left(\delta_{\alpha\beta}\delta_{\gamma\delta}+\delta_{\alpha\gamma}\delta_{\beta\delta}+\delta_{\alpha\delta}\delta_{\beta\gamma}\right)+\frac{1}{9}\delta_{\alpha\beta}\delta_{\gamma\delta},\end{eqnarray}
 whereas the correlator of two $g$ on different lattice sites is
zero, one obtains the important formula \begin{eqnarray}
 &  & \frac{1}{\mathcal{N}}\left\langle \left(\mathbf{a}\overleftrightarrow{\mathbf{g}}_{\mathbf{k}}\mathbf{b}\right)\left(\mathbf{c\overleftrightarrow{\mathbf{g}}_{-\mathbf{k}}d}\right)\right\rangle \notag\\
 & = & \frac{D_{R}^{2}}{15}\left\{ \left(\mathbf{a\cdot b}\right)\left(\mathbf{c\cdot d}\right)+\left(\mathbf{a\cdot c}\right)\left(\mathbf{b\cdot d}\right)+\left(\mathbf{a\cdot d}\right)\left(\mathbf{b\cdot c}\right)\right\} \notag\\
 &  & +\frac{D_{R}^{2}}{9}\left(\mathbf{a\cdot b}\right)\left(\mathbf{c\cdot d}\right).\label{RAAvr1}\end{eqnarray}
 In fact, the terms with $\left(\mathbf{a\cdot b}\right)\left(\mathbf{c\cdot d}\right)$
vanish in the expressions below. With the help of Eq. (\ref{RAAvr1})
one obtains \begin{eqnarray}
R_{\Vert} & = & -4m^{3}\frac{D_{R}^{2}}{15}\, Re\left[\frac{1}{\mathcal{N}}\sum_{\mathbf{k}}\frac{1}{\hbar}\int_{0}^{t}dt^{\prime}e^{-i\omega_{\mathrm{ex},\mathbf{k}}(t-t^{\prime})}\right.\notag\\
 &  & \times\left\{ \left(\mathbf{e}_{-}(t)\cdot\mathbf{e}_{+}(t^{\prime})\right)\left(\mathbf{n}(t)\cdot\mathbf{n}(t^{\prime})\right)\right.\notag\\
 &  & +\left.\left.\left(\mathbf{n}(t)\cdot\mathbf{e}_{+}(t^{\prime})\right)\left(\mathbf{e}_{-}(t)\cdot\mathbf{n}(t^{\prime})\right)\right\} \right].\label{RparIntegral}\end{eqnarray}
 After computer algebra using Eqs. (\ref{ne12tDep}) and (\ref{zetakDef})
one obtains \begin{eqnarray}
 &  & \left(\mathbf{e}_{-}(t)\cdot\mathbf{e}_{+}(t^{\prime})\right)\left(\mathbf{n}(t)\cdot\mathbf{n}(t^{\prime})\right)\notag\\
 &  & {}+\left(\mathbf{n}(t)\cdot\mathbf{e}_{+}(t^{\prime})\right)\left(\mathbf{e}_{-}(t)\cdot\mathbf{n}(t^{\prime})\right)\notag\\
 & = & 3x^{2}\left(1-x^{2}\right)\notag\\
 &  & {}+\frac{1}{2}\left(1-x\right)^{2}\left(1+2x\right)^{2}e^{i\omega_{h}\left(t-t^{\prime}\right)}\notag\\
 &  & {}+\frac{1}{2}\left(1+x\right)^{2}\left(1-2x\right)^{2}e^{-i\omega_{h}\left(t-t^{\prime}\right)}\notag\\
 &  & {}+\frac{1}{2}\left(1-x\right)^{2}\left(1-x^{2}\right)e^{2i\omega_{h}\left(t-t^{\prime}\right)}\notag\\
 &  & {}+\frac{1}{2}\left(1+x\right)^{2}\left(1-x^{2}\right)e^{-2i\omega_{h}\left(t-t^{\prime}\right)},\end{eqnarray}
 where $x$ is given by Eq. (\ref{xDef}) and $\omega_{h}$ by Eq.
(\ref{omegahDef}). Since in Eq. (\ref{RparIntegral}) $\omega_{\mathrm{ex},\mathbf{k}}>0,$
here one should keep only the resonant terms with $e^{i\omega_{h}\left(t-t^{\prime}\right)}$
and $e^{2i\omega_{h}\left(t-t^{\prime}\right)}$ that satisfy the
energy conservation and cause transitions. One obtains \begin{eqnarray}
R_{\Vert} & = & -\frac{4m^{3}D_{R}^{2}}{15\hbar}\frac{1}{\mathcal{N}}\sum_{\mathbf{k}}\notag\\
 &  & \times\left\{ \frac{1}{2}\left(1-x\right)^{2}\left(1+2x\right)^{2}\pi\delta\left(\omega_{\mathrm{ex},\mathbf{k}}-\omega_{h}\right)\right.\notag\\
 &  & +\left.\frac{1}{2}\left(1-x\right)^{2}\left(1-x^{2}\right)\pi\delta\left(\omega_{\mathrm{ex},\mathbf{k}}-2\omega_{h}\right)\right\} \notag\\
 & = & -\frac{2m^{3}D_{R}^{2}}{15\hbar}\left(1-x\right)^{2}\left[\left(1+2x\right)^{2}\pi\rho_{\mathrm{ex}}(\omega_{h})\right.\notag\\
 &  & \qquad+\left.\left(1-x^{2}\right)\pi\rho_{\mathrm{ex}}(2\omega_{h})\right].\end{eqnarray}
 With the help of Eq. (\ref{rhoeps}) one obtains \begin{equation}
R_{\Vert}=-\frac{2m^{3/2}}{15\pi}\frac{D_{R}^{2}}{J}\sqrt{\frac{h}{J}}\Phi_{\Vert}(x),\end{equation}
 where $\Phi_{\Vert}(x)$ is given by Eq. (\ref{PhiPar}).

For $R_{2}$ from Eqs. (\ref{RCompsFourier}) and (\ref{zetakSol})
one obtains \begin{eqnarray}
R_{\bot} & = & 4m^{3}\, Re\left[\frac{1}{\mathcal{N}^{2}}\sum_{\mathbf{k}}\frac{i}{\hbar}\int_{0}^{t}dt^{\prime}e^{-i\omega_{\mathrm{ex},\mathbf{k}}(t-t^{\prime})}\right.\notag\\
 &  & \times\left\{ \left(\mathbf{e}_{1}(t)\overleftrightarrow{\mathbf{g}}_{-\mathbf{k}}\mathbf{e}_{-}(t)\right)-\left(\mathbf{n}(t)\overleftrightarrow{\mathbf{g}}_{-\mathbf{k}}\mathbf{n}(t)\right)\right\} \notag\\
 &  & \times\left.\left(\mathbf{e}_{+}(t^{\prime})\overleftrightarrow{\mathbf{g}}_{\mathbf{k}}\mathbf{n}(t^{\prime})\right)\right]\notag\\
 & = & \frac{4m^{3}D_{R}^{2}}{15}\, Re\left[\frac{1}{\mathcal{N}}\sum_{\mathbf{k}}\frac{i}{\hbar}\int_{0}^{t}dt^{\prime}e^{-i\omega_{\mathrm{ex},\mathbf{k}}(t-t^{\prime})}\right.\notag\\
 &  & \times\left\{ \left(\mathbf{e}_{1}(t)\cdot\mathbf{e}_{+}(t^{\prime})\right)\left(\mathbf{e}_{-}(t)\cdot\mathbf{n}(t^{\prime})\right)\right.\notag\\
 &  & {}+\left(\mathbf{e}_{1}(t)\cdot\mathbf{n}(t^{\prime})\right)\left(\mathbf{e}_{-}(t)\cdot\mathbf{e}_{+}(t^{\prime})\right)\notag\\
 &  & -\left.\left.2\left(\mathbf{n}(t)\cdot\mathbf{e}_{+}(t^{\prime})\right)\left(\mathbf{n}(t)\cdot\mathbf{n}(t^{\prime})\right)\right\} \right].\end{eqnarray}
 Computer algebra yields \begin{eqnarray}
 &  & \frac{i}{\sqrt{1-x^{2}}}\left\{ \left(\mathbf{e}_{1}(t)\cdot\mathbf{e}_{+}(t^{\prime})\right)\left(\mathbf{e}_{-}(t)\cdot\mathbf{n}(t^{\prime})\right)\right.\notag\\
 &  & +\left(\mathbf{e}_{1}(t)\cdot\mathbf{n}(t^{\prime})\right)\left(\mathbf{e}_{-}(t)\cdot\mathbf{e}_{+}(t^{\prime})\right)\notag\\
 &  & -\left.\left.2\left(\mathbf{n}(t)\cdot\mathbf{e}_{+}(t^{\prime})\right)\left(\mathbf{n}(t)\cdot\mathbf{n}(t^{\prime})\right)\right\} \right]\notag\\
 & = & -3x^{3}-\frac{1}{2}(1-x)(1+2x)^{2}e^{i\omega_{h}\left(t-t^{\prime}\right)}\notag\\
 &  & +\frac{1}{2}(1+x)(1-2x)^{2}e^{-i\omega_{h}\left(t-t^{\prime}\right)}\notag\\
 &  & -\frac{1}{2}(1-x)^{2}(2+x)e^{2i\omega_{h}\left(t-t^{\prime}\right)}\notag\\
 &  & +\frac{1}{2}(1+x)^{2}(2-x)e^{-2i\omega_{h}\left(t-t^{\prime}\right)}.\end{eqnarray}
 This results in \begin{eqnarray}
R_{\bot} & = & -\frac{2m^{3}D_{R}^{2}}{15\hbar}\sqrt{1-x^{2}}(1-x)\pi\left\{ (1+2x)^{2}\rho_{\mathrm{ex}}(\omega_{h})\right.\notag\\
 &  & \qquad\times\left.(1-x)(2+x)\rho_{\mathrm{ex}}(2\omega_{h})\right\} .\end{eqnarray}
 Finally, using Eq. (\ref{rhoeps}) again, this expression can be
brought into the form \begin{equation}
R_{\bot}=-\frac{m^{3/2}}{5\pi}\frac{D_{R}^{2}}{J}\sqrt{\frac{h}{J}}\sqrt{1-x^{2}}\Phi_{\bot}(x),\end{equation}
 where $\Phi_{\bot}(x)$ is given by Eq. (\ref{PhiPerp}). Finally,
with \begin{equation}
\Gamma_{\Vert}(x)=-\frac{1}{\hbar}R_{\Vert}(x),\qquad\Gamma_{\bot}(x)=-\frac{1}{\hbar}\frac{R_{\bot}(x)}{\sqrt{1-x^{2}}}\label{GammaRes}\end{equation}
 one obtains Eq. (\ref{LLBSWEq}).

\bibliographystyle{apsrev}
\bibliography{hkbib}
 %\bibliography{hkbib, gar-own,gar-tunneling,gar-relaxation,gar-oldworks,gar-books,gar-general,gar-surface-nano,gar-superparamagnetic}

\end{document}